\documentclass[aps,prx,reprint,superscriptaddress,nofootinbib]{revtex4-2}
\usepackage{amsmath,amssymb,mathtools,bm}
\usepackage{graphicx,booktabs,xcolor,microtype,tikz}
\usetikzlibrary{arrows.meta,positioning,shapes.geometric,calc}
\usepackage[colorlinks=true,allcolors=blue!55!black]{hyperref}
\usepackage[nameinlink,capitalise]{cleveref}
\usepackage{comment}
\definecolor{qblue}{HTML}{2864A8}
\definecolor{qteal}{HTML}{168579}
\definecolor{qred}{HTML}{B84343}
\definecolor{qgold}{HTML}{C88918}
\definecolor{qpurple}{HTML}{70549A}
\definecolor{qgray}{HTML}{56616D}

\newcommand{\Ftwo}{\mathbb F_2}
\newcommand{\id}{\mathbb I}

\newcommand{\im}{\operatorname{im}}
\newcommand{\normop}[1]{\left\lVert #1\right\rVert}
\newcommand{\codeH}{H_{\mathrm{code}}}
\newcommand{\effH}{H_{\mathrm{eff}}}
\newcommand{\leanname}[1]{\nolinkurl{#1}}
\newtheorem{theorem}{Theorem}

\newcommand{\addAA}[1]{\textcolor{black}{#1}}
\providecommand{\khz}{\ensuremath{\,\mathrm{kHz}}}
\providecommand{\ghz}{\ensuremath{\,\mathrm{GHz}}}
\providecommand{\ns}{\ensuremath{\,\mathrm{ns}}}

\begin{document}

\title{Agentic Pipelines for Mapping Topological Codes to Quantum Hardware}
\author{Arpit Arora}
\affiliation{NVIDIA, Santa Clara, California 95051, USA}
\author{Adam Holmes}
\affiliation{NVIDIA, Santa Clara, California 95051, USA}
\author{Krysta M. Svore}
\affiliation{NVIDIA, Santa Clara, California 95051, USA}
\author{Taylor L. Patti}
\affiliation{NVIDIA, Santa Clara, California 95051, USA}
\date{September 1, 2026}

\begin{abstract}
Mapping a prescribed many-body Hamiltonian to a particular quantum hardware is generally an approximate process, resulting in both intended and unwanted quantum dynamics. We introduce an AI-agent-assisted workflow that uses Lean to establish whether an effective device Hamiltonian retains the intended low-energy subspace of the target many-body Hamiltonian. We demonstrate this approach with topological quantum error correction (QEC), providing, to our knowledge, the first Lean-verified pipeline connecting a code space to low-energy states of a device model. Starting from qubits and parity checks, the pipeline constructs the ideal Hamiltonian whose lowest energy states form the intended code space, and determines the excitation gap that protects that code space. It then tests if the device retains the required number of isolated low-energy states and quantifies how closely their subspace matches the intended code space. For this test we compare three bounds on energy shifts by unwanted interactions, namely those obtained by: summing the interaction strengths, grouping the anticommuting Pauli interactions, and diagonalizing the Hamiltonian. We find that Pauli grouping yields a bound that is tighter than the conventional sum and closer to the diagonalization result, providing a method that avoids overly conservative compensation requirements in hardware design. The workflow also identifies the strongest unwanted interactions and the qubits they affect, providing specific targets for improving device couplings and controls. We demonstrate our framework using an eight-qubit toric-code array on a superconducting device, where the same driven device generates both the required four-qubit interactions as well as unwanted dynamics. An AI agent identifies viable device parameters within a prescribed leakage budget, and Lean verifies four isolated low-energy states close to the ideal code space. Analysis of the residual interactions identifies the dominant unwanted couplings. Our workflow provides a reusable tool for the analog-compilation of a QEC code on a quantum device, which allows for predictive engineering of quantum hardware development.
\end{abstract}

\maketitle

\section{Introduction}
\label{sec:intro}

Quantum simulation seeks to realize many-body Hamiltonians whose dynamics and low-energy phases are difficult to observe through traditional experiments or study numerically~\cite{Georgescu2014,Daley2022}. Whereas digital simulation decomposes these Hamiltonians into discrete quantum gate sequences, analog simulation engineers them directly from continuous native quantum interactions, enabling continuous access to their spectra and dynamics fo quantum states~\cite{Altman2021}. This turns hardware design into an analog-compilation problem: given a target Hamiltonian, determine the device architecture and control sequences that best realize the target effective Hamiltonian. 

Topological quantum error correction (QEC) provides a particularly stringent example~\cite{Kitaev2003,Dennis2002}. Its stabilizer Hamiltonian must not only produce selected interaction strengths, but also an encoded low-energy sector with the correct degeneracy, excitation gap, and logical structure \cite{Satzinger2021,Semeghini2021,Sameti2017,Wilkinson2020,Menke2021,BravyiHastingsMichalakis2010,Hangleiter2017,Shaffer2021}. Gate- and measurement-based approaches have demonstrated topological-code
state preparation, logical operations, and error suppression across trapped-ion, neutral-atom, and superconducting platforms~\cite{Nigg2014,Bluvstein2026,GoogleQAI2025}. Quantum-simulation experiments seek to embed these codes directly in interacting quantum matter through engineered multiqubit interactions for the preparation and detection of topologically ordered states~\cite{Sameti2017,Dai2017,Verresen2021,Semeghini2021}. The simulation of topological QEC codes on a device raises a fundamental question: \addAA{how can we ensure that the desired code Hamiltonian emerges from the collective many-body interactions of the device?}

Previous works have addressed components of this broader problem, but they have not produced an end-to-end solution that maps code specifications to target device \addAA{designs}. 
Stabilizer and homological-code theory determine the logical structure and error-detection properties of an ideal code~\cite{Calderbank1996,Steane1996,Dennis2002}. On the hardware side, driven superconducting circuits have been proposed that generate toric-code interactions~\cite{Sameti2017}, with automated design of four-qubit couplers~\cite{Menke2021}. Circuit-analysis tools construct Hamiltonians
from circuit descriptions \cite{QuCAT2020,SQcircuit2023}, while Hamiltonian-learning methods infer interactions from measurements~\cite{Bairey2019}.

\addAA{Connecting Hamiltonian simulation to device specifications requires coordinating physical modeling, mathematical derivations, and numerical simulations across multiple tools. AI-assisted workflows can support this process by orchestrating and perfomring multistep tasks. Recently, mathematical arguments developed by AI can be expressed as formal proofs in Lean, a proof-checking program, which independently checks whether the conclusions follow from the stated assumptions~\cite{AxProver2025,MerLean2026,Lu2026FormalFlow}.}
Recent applications \addAA{of Lean in the quantum domain} include information-theoretic limits~\cite{SteinLean2025,LeanQuantum2026,LeanQIT2026}, algorithm analysis and optimization~\cite{ShorLean2026,Kol2026}, error-correction codes~\cite{LeanQEC}, automated proof generation~\cite{AxProver2025,MerLean2026,Lu2026FormalFlow,AxQM2026}, and Hamiltonian compilation~\cite{VerifiedCompiler2025}. Complementary approaches use independent computational evaluators to guide LLM searches toward quantum circuits with reduced Toffoli-gate and qubit requirements~\cite{Gu2026CircuitDiscovery}.

\begin{figure*}
    \centering
    \includegraphics[width=0.9\linewidth]{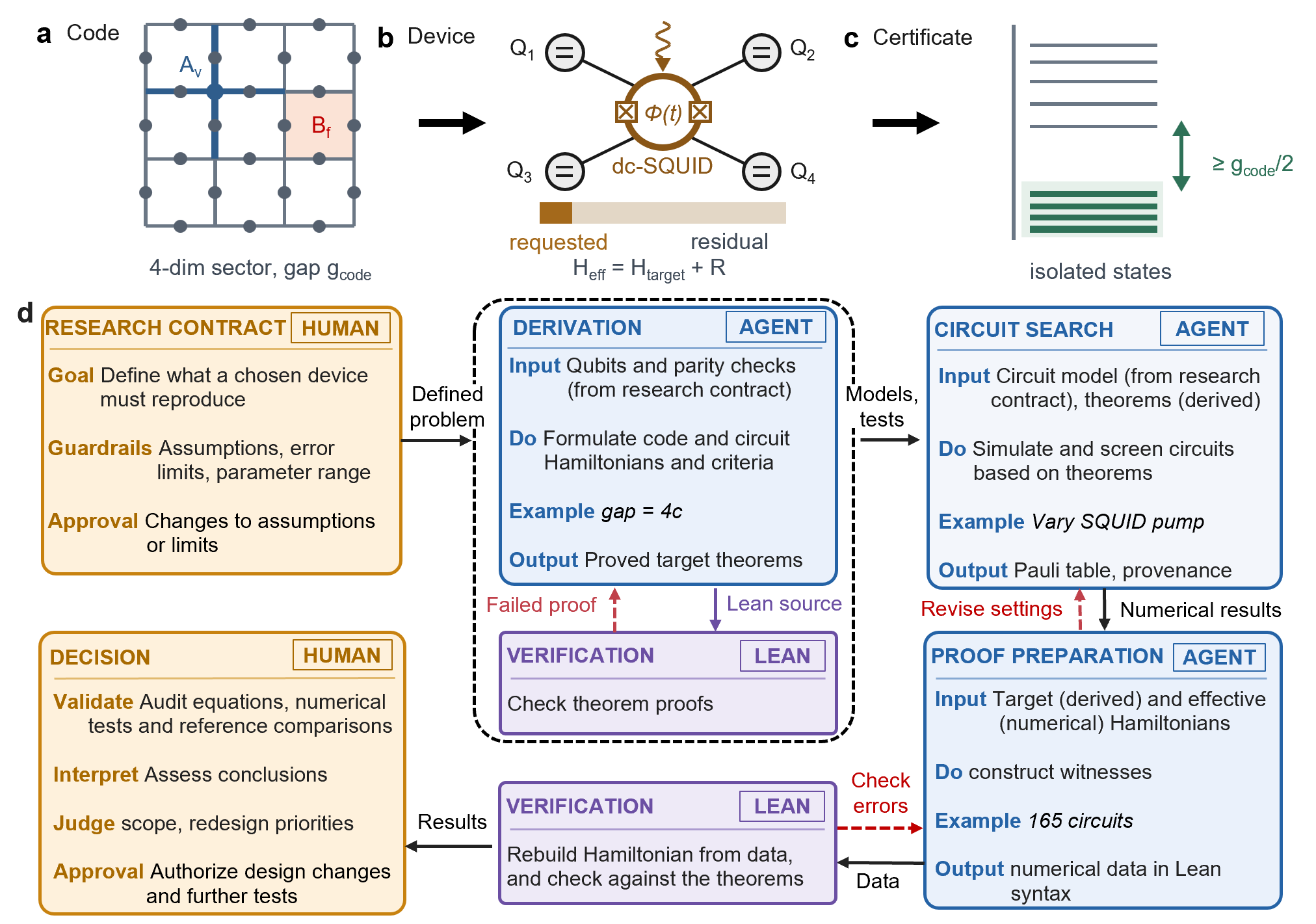}
    \caption{\textbf{From a topological-code request to a certified physical
low-energy sector.}
(a) Cell-complex incidence data specify the periodic toric-code
Hamiltonian, its four-dimensional code sector, and its gap.
(b) A driven dc-SQUID connected to an array of four qubits ($Q_{1.2.3.4}$) generates the requested four-body checks together
with thousands of non-target residual terms, producing an effective Hamiltonian approximating the QEC code.
(c) After alignment with the target Hamiltonian, Lean verifies a rigorous bound on the
residual operator. The resulting spectral corridor contains exactly four
isolated low-energy states, as well as a detailed profile of residual channels. (d) Agentic workflow employed in this work that starts with a human defined research contract, which is then decomposed by the agent, which in turn performs the corresponding mathematical and numerical calculations. The calculations are verified and certified under a Lean diagnostic loop. The human ultimately evaluates the correctness and physical interpretation of the workflow output.
}
    \label{fig1}
\end{figure*}

Building on these advances in agent-guided discovery and formal verification, we \addAA{propose a Lean verified workflow that starts from a target QEC Hamiltonian, searches the parameters of a specified analog device architecture, and uses Lean to verify whether the resulting effective device Hamiltonian supports the target QEC Hamiltonian to a given tolerance. Starting from the definition of the QEC code, rather than a device layout, we construct the target Hamiltonian from the qubits and parity checks specified by that code. }
We establish conditions that guarantee the device \addAA{design} retains the target number of low-energy states, and quantify both their separation from higher-energy states and how far the subspace they span differs from the intended code space. We resolve the device Hamiltonian into Pauli interactions and subtract the target Hamiltonian to identify the residual. Our certification criterion requires a verified upper bound on the residual’s norm to lie below half the target excitation gap. We group anticommuting Pauli interactions as they cannot all reinforce one another fully in the same quantum state. In particular, the agent proposed this approach in response to a human request for a tighter residual bound, and as a result anticommuting-group methods developed for variational quantum algorithms~\cite{Izmaylov2020,Ralli2021} were adapted, to our knowledge for the first time, to QEC device Hamiltonians. This application enabled a tighter bound on the energy shifts caused by the residual channels than the conventional estimate, which simply adds the magnitude of their interaction strengths. We demonstrate the pipeline on an eight-qubit model of the \(L=2\) toric code, using a driven superconducting architecture~\cite{Kitaev2003,Sameti2017} \addAA{in which four-qubit interactions are generated by driving nonlinear SQUID couplers. The same drive also produces unwanted interactions, raising the challenge of how to strengthen the desired interactions while preserving the low-energy sector required by the target Hamiltonian}. 
Our workflow can also be deployed before fabrication to reject or refine couplers and controls.

\section{Topological QEC Hamiltonian and device engineering}
\label{sec:map}
In topological QEC codes, such as toric and surface codes, qubits are arranged on a lattice, and the information that they encode is spread across a group of qubits to protect it from local perturbations~\cite{terhal2015quantum}. As illustrated in Fig.~\ref{fig1}a, this ensemble of qubits is monitored through local checks called stabilizers composed of vertex ($A_v$) and plaquette ($B_f$) operators. When these checks are violated, an excitation is created in the system. In terms of an analog hardware, a topological QEC Hamiltonian can be realized on \addAA{an} underlying device if the the code space (the set of states encoding the information) is separated from the rest of the system and noise channels by an excitation gap~\cite{Sameti2017}. 

We consider the toric code on an \(L\times L\) periodic square lattice,
with \(L\geq2\), placing one physical qubit on each edge. The star operator \(A_v\) applies Pauli \(X\) to the four qubits meeting at vertex \(v\), while the plaquette operator \(B_f\) applies Pauli \(Z\) to the four qubits surrounding face \(f\). These are therefore \(XXXX\) and \(ZZZZ\) interactions on their respective groups of qubits, with the identity acting on all remaining qubits. The star and plaquette checks commute, so measuring one does not disturb the outcomes of the others. Each check returns either \(+1\), indicating that its local requirement is satisfied (no error detected), or \(-1\), indicating a violation (an error signal). The code space contains the states satisfying every check~\cite{Kitaev2003,Dennis2002,bombin2013introduction}.

Next, with \(J_X\) and \(J_Z\) as the respective costs of violating a star or plaquette check, we write the standard Hamiltonian~\cite{Kitaev2003,Dennis2002} (see the Supplmental Material (SM) \cref{sm:sec-code} for homological construction)
\begin{equation}
 \codeH=
 J_X\sum_{v\in V}\frac{\id-A_v}{2}
 +J_Z\sum_{f\in F}\frac{\id-B_f}{2},
 \label{eq:code-hamiltonian}
\end{equation}
with $J_X,J_Z>0$. Here \(V\) and \(F\) denote the vertices and faces, and \(\id\) is the identity operator. After removing the constant energy offset, each requested \(XXXX\) interaction has coefficient \(-J_X/2\), and each \(ZZZZ\) interaction has coefficient \(-J_Z/2\). These are the effective four-qubit coupling strengths the hardware must generate. The values of \(J_X\) and \(J_Z\) set the energy cost of violating a check, without changing the desired code states. 
For example, starting from a state satisfying every check, a Pauli \(Z\)-error on one edge changes the two star checks at its endpoints from \(+1\) to \(-1\). Each affected term in Eq.~\ref{eq:code-hamiltonian} then contributes \(J_X\), placing the altered state \(2J_X\) above the code states. As the error changes the state, \(J_X\) sets its energy penalty, and increasing this penalty does not itself introduce or correct the error. Similarly, a Pauli-\(X\) error violates the two adjacent plaquette checks and costs \(2J_Z\). We note that, because each qubit belongs to two checks of each type, a Pauli error flips the corresponding check outcomes in pairs. The lowest excitation therefore requires two violated checks, giving an ideal gap of \(2\min(J_X,J_Z)\)~\cite{Kitaev2003,Dennis2002, YangBuecheleTrivedi2025}.

For the toric code on this periodic lattice, topology fixes the ideal ground space to contain four independent states and their superpositions~\cite{Kitaev2003,Dennis2002}. The hardware must therefore reproduce a four-dimensional low-energy space, not merely one low-energy state.  Besides these universal features, the lattice size controls the smallest undetectable error: at least \(L\) qubits on $L\times L$ lattice must be affected to change a code state without violating any check. Thus the toric-code family has distance \(d=L\). The origin of the checks and $H_{\rm code}$ from the lattice geometry and principles of homology is described in the SM \cref{sm:sec-code} and is formalized in Lean (including Hermicity, Pauli algebra, commutation of local operators, excitation constraints and code distance).

In an active QEC implementation, each
four-body check can be measured through a sequence of one- and two-qubit gates. Conversely, in an analog implementation, such as that considered here, the device must generate the check itself as a term in its effective Hamiltonian. We set in Eq.~\ref{eq:code-hamiltonian} \(J_X=J_Z=2c\) such that requested code Hamiltonian is 
\begin{equation}
 H_0(L)=H_{\rm code}(J_X=J_Z =2c)
 \label{eq:target}
\end{equation}
where \(c\) is the stabilizer coupling strength, the common magnitude of the requested four-qubit \(XXXX\) and \(ZZZZ\) interactions. A violated check costs \(2c\), and 
the smallest ideal excitation costs
\begin{equation}
 g_{\rm code}=4c .
 \label{eq:ideal-gap}
\end{equation}
Thus \(c\) controls both the
strength of the engineered stabilizer interaction and the ideal spectral protection available against unwanted coherent terms.

Several platforms offer routes to the required multiqubit interactions, with their own engineering constraints. Optical-lattice experiments have demonstrated four-spin exchange by suppressing the otherwise dominating lower-order processes~\cite{Dai2017}. In Rydberg arrays, control laser excitation of atoms can prevent nearest neighbor exchange, allowing preparation of topological states~\cite{Verresen2021,Semeghini2021}. However, preparing a topological state does not establish that it is a ground state of the device as, interactions from distinct atoms can favor a different non-topological ground state~\cite{Semeghini2021}.

Superconducting circuits offer another route to realizing topological states, using nonlinear elements, tunable couplings, and microwave control to engineer the required multiqubit interactions~\cite{Arora2026QuantumLaboratories}.
Driven dc-SQUID couplers~\cite{Sameti2017} and flux-qubit couplers~\cite{Menke2022,Menke2021} have demonstrated tunable three- and four-body interactions. Coupled Josephson parametric oscillators offer another source of four-body mechanism for quantum annealing~\cite{Puri2017,Miyazaki2025}. However, all superconducting devices face a common problem wherein the desired four-body interactions are generated through the same Josephson nonlinearity that also produces lower-body interactions~\cite{Sameti2017,Wilkinson2020}.  Increasing the desired coupling \(c\) can therefore also strengthen unwanted interactions or increase leakage out of the computational
subspace. A large target coefficient alone does not establish that the device realizes \(H_0\). This tension between strengthening the desired four-body checks and controlling the accompanying
Hamiltonian is the physical problem addressed in this work. For our demonstration, we select the driven dc-SQUID coupler coupled to four qubits~\cite{Sameti2017}, see Fig.~\ref{fig1}b. This was specifically proposed to generate the toric code's four-body device checks. For this architecture, \addAA{realizing the requested Hamiltonian requires choosing}
experimentally accessible coupler parameters, pump frequencies, amplitudes, and phases.

\section{Agentic Derivation and Verification Workflow}
\label{sec:agentic-workflow}
We follow an AI agent-guided workflow governed by a human-\addAA{specified} research contract. This \addAA{contract} specifies the scientific objective, permitted assumptions, computational constraints, and requirements for accepting a result. 
In a multi-agent implementation, specialized agents use different tools to develop mathematical arguments and perform calculations, while humans review results and approve changes to the scientific scope and assumptions. Mathematical statements and their proofs are written in Lean 4, a programming language and proof assistant whose proof-checking kernel verifies that the supplied proofs establish the stated conclusions under their declared assumptions \cite{deMouraUllrich2021}. The execution sandbox restricts which files and tools the agents
may access. Neither sandboxing nor proof checking automatically validates the simulation programs or establishes that the assumed physics describes the input problem. A successful Lean proof cannot override an unresolved numerical discrepancy or justify an unsupported physical assumption. We address these concerns through human-directed quantitative validation: auditing the implemented equations, numerical convergence, error budgets, and comparisons with reference calculations, and requiring discrepancies to be resolved before accepting conclusions. 

We organize this workflow (see Fig.~\ref{fig1}d) around the central question: does a proposed \addAA{device design} reproduce the low-energy sector requested by a topological code? The human provided research contract records the device model, accuracy requirements, and intended conclusions, evolving as the study progresses (see research contract box in Fig.~\ref{fig1}d). From the qubit arrangement and stabilizer checks, the agent constructs the ideal Hamiltonian and formalizes its properties  (see derivation box in Fig.~\ref{fig1}d). It then derives and formalizes \cref{thm:criterion} \cref{sec:criterion}), the known condition that relates the excitation of gap of the target Hamiltonian (see Eqs.~\ref{eq:target} and \ref{eq:ideal-gap}) to the norm of residual interactions in the effective device Hamiltonian. Subsequently, the agent was asked to find a tighter bound for \cref{thm:criterion}, and proposed adapting anticommuntation-group methods from variational quantum algorithms~\cite{Izmaylov2020,Ralli2021}. The bound was derived and formalized in Lean, and presented as \cref{thm:anticommuting-bound} in \cref{sec:criterion}. The agent-Loop feedback for derivation and verification is shown in dashed box in \addAA{the} Fig.~\ref{fig1}d.

The agent then applies these theorems to numerical simulations of the device Hamiltonian, searching across circuit parameters to identify settings that satisfy the required conditions. The numerical results were translated to Lean syntax by the agent. Lean then reconstructs the device Hamiltonian from these coefficients, compares it with the independently specified target, and checks the proofs used to apply the theorems. Failed proof checks return errors for the agent to resolve, while bounds that do not satisfy the theorem motivate further circuit searches. The agent-Loop feedback for circuit search, numerical calculation and Lean verification is shown in Fig.~\ref{fig1}d. 

OpenAI Codex agents (GPT-5-family models) developed mathematical
derivations, wrote and executed simulation and parameter-search
programs, and wrote and revised Lean proofs
~\cite{OpenAICodexGuide2026,OpenAIGPT5}, while Anthropic Claude
(Opus 5)~\cite{AnthropicOpus5} provided external critical review.
Human authors set the scientific scope, and interpret the results. As such, they retain responsibility for all claims. 

\section{From a compiled Hamiltonian to a finite-error code certificate}
\label{sec:criterion}
In this section, we discuss how we deduce the conditions under which the target Hamiltonian, $H_0$, can be realized on an underlying quantum device. For the Hamiltonian \(H_0\), we define ground projector, \(P_0=\sum_\alpha|\psi_\alpha\rangle\langle\psi_\alpha|\), where the states \(|\psi_\alpha\rangle\) form an orthonormal basis of the ground state subspace of $H_0$. This projector identifies the target subspace whose proximity to the device’s low-energy subspace is bounded in \cref{thm:criterion} (see \cref{eq:projector-bound} below) and evaluated numerically in our demonstration. The corresponding ground space has dimension \(m=\operatorname{rank}(P_0)\), excitation gap \(g_{\rm code}\), and stabilizer scale $c$. The device and code descriptions will, in general, not be fully compatible, prompting us to define the effective Hamiltonian $\effH$ in terms of unitary rotations $U$ and its additive residual $R$
\begin{equation}
 \effH=U H_0U^\dagger+R.
 \label{eq:normal-form}
\end{equation}
More specifically, a device may generate interactions along different qubit axes from those used to specify the target. We account for this possibility by selecting \(U\) from a predefined finite set of rotations acting separately on individual qubits, such that each rotation applies consistently to every interaction involving that qubit. Thus \(UH_0U^\dagger\) retains the target interaction while allowing these specified changes of axes. The residual \(R\) contains the difference between the device and rotated target Hamiltonians,  unwanted interactions, and errors in the strengths or signs of the requested interactions.

If a central model calculates $\normop{R}$, the numerical backend bounds \(\normop{R}\le r_{\rm cert}\), so that Lean verification of numerical Hamiltonian does not depend on floating-point norm estimates. Additionally, we introduce a tolerance parameter $\eta$, which accounts for any numerical uncertainty due to truncated Hilbert space dimensions and finite time steps, as well as for the conversion of numerical decimals to rational numbers, which is often necessary because Lean treats fractions exactly. Consequently, \(r=r_{\rm cert}+\eta\) is a rigorous upper bound on the total mismatch. The following theorem shows why this bound matters, demonstrating that if the total mismatch is smaller than half of the target excitation gap, the device retains the requested number of isolated low-energy states.
\begin{theorem}[Finite encoded-sector criterion]
\label{thm:criterion}
Let \(H_0\) and \(\effH\) be Hermitian Hamiltonians on the same
finite-dimensional Hilbert space. Suppose \(H_0\) has an
\(m\)-dimensional zero-energy ground space with projector \(P_0\)
and excitation gap of at least \(g_{\rm code}>0\).
Assume Eq.~\eqref{eq:normal-form} holds with \(U\) unitary and
\(\normop{R}\le r\). If
\begin{equation}
 2r<g_{\rm code},
 \label{eq:criterion}
\end{equation}
then \(\effH\) has exactly \(m\) eigenvalues below
\(g_{\rm code}/2\), separated from its remaining spectrum by
at least \(g_{\rm code}-2r\). The spectral projector
\(P_{\rm low}\) for $m$ lowest energy states of $H_{\rm eff}$ satisfies
\begin{equation}
 \normop{P_{\rm low}-UP_0U^\dagger}
 \le \frac{r}{g_{\rm code}-r}.
 \label{eq:projector-bound}
\end{equation}
\end{theorem}
Theorem~\ref{thm:criterion} states an established criterion
\cite{Bhatia1997,DavisKahan1970,Bravyi2010,KnillLaflamme1997}.
Our workflow (see Fig.~\ref{fig1}d) formalizes it in Lean and checks whether the calculated device Hamiltonian satisfies its conditions.

Physically, \(R\) broadens the ideal code manifold into a low-energy band and
pulls the excited manifold toward it.  A norm-\(r\) perturbation moves either
edge by at most \(r\), such that if the broadened intervals do not meet, the physical
Hamiltonian retains exactly the \(m\) states delineated by the ideal code.  The
projector estimate additionally identifies their span with a deformation of
the intended code space.  For example, \(r=g_{\rm code}/4\) guarantees a
remaining separation of at least \(g_{\rm code}/2\) and a projector distance
of at most \(1/3\), also see SM \cref{sm:sec-criterion}. Theorem~\ref{thm:criterion} is a sufficient, but not necessary, condition. For instance, Eq.~\ref{eq:criterion} can be violated by a term commuting with every stabilizer that shifts energies without mixing the code states and excited states.

To construct \(R\), both the compiled Hamiltonian and the target are written in the \(n\)-qubit Pauli basis. After the target terms are removed, the remaining operator is \(R=\sum_a r_aP_a\), where \(a\) labels a retained channel and \(r_a\in\mathbb R\) is its coupling strength. Importantly, the Pauli word \(P_a\in\{\id,X,Y,Z\}^{\otimes n}\) is a tensor product specifying one physical interaction channel, such as a single-qubit field \(X_i\), a two-qubit coupling \(Z_iZ_j\), or a four-qubit check \(X_iX_jX_kX_l\). According to the conventionally used bound, the operator norm satisfies 
\begin{equation}
\label{eq:coefficientwise}
\normop{R}\le\sum_a|r_a|,
\end{equation}
obtained by applying the triangle inequality term by term, using
\(\normop{P_a}=1\) for every Pauli word. Such coefficient-one-norm bounds are
widely used in Hamiltonian simulation, truncation-error analysis, and
Hamiltonian certification
\cite{Childs2021,Loaiza2023,Fuller2026}. These bounds can be highly conservative because they
require the Pauli terms to admit a common state that simultaneously realizes
their extremal contributions, a limitation made explicit in recent analyses of
Pauli truncation error~\cite{Fuller2026} (for more details, see the SM \cref{sm:sec-bounds}). Thus, we focus on the commutation relations among the
residual channels to extract the missing physical information. To meet this human-defined objective, the agent proposed the norm identity for anticommuting Pauli groups:

\begin{theorem}[Commutation-aware Pauli bound]
\label{thm:anticommuting-bound}
Partition all channel labels into groups \(\mathcal G_j\) such that any two
distinct Pauli words within one group anticommute,
\(P_aP_b=-P_bP_a\).  Then
\begin{align}
 \normop{\sum_{a\in\mathcal G_j}r_aP_a}
   &=\sqrt{\sum_{a\in\mathcal G_j}r_a^2},\nonumber\\
 \normop{R}
   &\le\sum_j\sqrt{\sum_{a\in\mathcal G_j}r_a^2}.
 \label{eq:anticommuting}
\end{align}
\end{theorem}
\cref{thm:anticommuting-bound} was proved and verified using the workflow in Fig.~\ref{fig1}d, and its analytical proof is provided in the SM \cref{sm:sec-bounds}. Physically, \cref{thm:anticommuting-bound} implies that anticommuting channels cannot reinforce one another on the same quantum state,
so their strengths add in quadrature within each group.  Consequently, \(N\) anticommuting channels of
comparable magnitude \(\epsilon\) contribute \(\epsilon\sqrt N\), instead of
the \(N\epsilon\) assigned by coefficientwise accounting.  This algebraic
identity underlies unitary-partitioning methods in variational quantum
algorithms \cite{Izmaylov2020,Ralli2021}. Here, we convert it into a certified
error budget for compiled QEC Hamiltonians. In our pipeline, an agent proposes the partition, and Lean verifies that
every residual term is included and that every claimed anticommutator
vanishes. Restrictions on finite anticommuting Pauli families are discussed
in Ref.~\cite{Sarkar2021}.

The grouping of Pauli words in \cref{eq:coefficientwise} and \cref{eq:anticommuting} give a bound over $\normop{R}$ through coefficients of Pauli words without forming the full $R$ matrix. For a small finite device, we can construct the complete \(2^n\times2^n\) computational-basis matrix of
\(R=\sum_a r_a P_a\), and formulate a tighter bound for \cref{thm:criterion}. We define
\begin{equation}
\label{eq:fullcheck}
 A_{{\rm cert}}=r_{\rm cert}^2\id-R^2,
\end{equation}
where $A_{{\rm cert}}$ is an auxiliary matrix that tests if energy shifts due to $R$ lie between $-r_{\rm cert}$ and $r_{\rm cert}$. Because \(R\) is Hermitian, \(A_{{\rm cert}}\succeq0\) implies that every eigenvalue
\(\lambda\) of \(R\) satisfies \(|\lambda|\le{r_{\rm cert}}\), and therefore
\(\normop{R}\le{r_{\rm cert}}\). To certify this, a standard numerical-proposal--exact-verification strategy~\cite{Rump2006,PeyrlParrilo2008,Dostert2021} is followed in Lean. The exact witness obtained by diagonalizing the full matrix, however, is computationally expensive, as it retains all interference between Pauli channels. Computing all three bounds allows us to compare their tightness against their computational demands. For larger systems, where using \cref{eq:fullcheck} becomes impractical, bounds calculated directly from the Pauli interactions provide a practical alternative. As shown in \cref{sec:squid}, \cref{thm:anticommuting-bound} substantially improves on the Eq.~\ref{eq:coefficientwise} estimate and approaches the bound imposed by Eq.~\ref{eq:fullcheck}. This supports its use for larger systems with manageable Pauli inventories, without constructing the full Hamiltonian matrix.

The resulting pipeline is independent of the code family and hardware
mechanism.  The code specification supplies
\((H_0,P_0,m,g_{\rm code})\), while a circuit model supplies \(\effH\). Eqs.~\ref{eq:ideal-gap},\ref{eq:anticommuting},\ref{eq:fullcheck} then bound the residual \(R\), and
\cref{thm:criterion} determines the dimension, spectral
isolation, and projector error of the implemented low-energy sector.  The remainder of the paper demonstrates the method for a driven-SQUID realization of the toric-code Hamiltonian.

\section{Demonstration with a driven-SQUID coupler}
\label{sec:squid}
\subsection{Single cell demonstration}
We apply our workflow to the flux-driven dc-SQUID architecture~\cite{Sameti2017}, where four qubits with distinct transition frequencies couple through two intermediate nodes with phases $\varphi_a$ and $\varphi_b$, respectively. The single cell Hamiltonian is
\begin{equation}
 H_{\rm cell}(t)=H_{\rm lin}
 -2E_J
 \cos\!\left[\frac{\varphi_{\rm dc}+\varphi_d(t)}{2}\right]
 \cos\widehat{\varphi},
 \label{eq:cell}
\end{equation}
where \(E_J\) is the Josephson energy of each junction and \(\widehat{\varphi}\) is the phase difference across the SQUID. The external flux is expressed as a dimensionless phase, \(\varphi_{\rm ext}=2\pi\Phi_{\rm ext}/\Phi_0 =\varphi_{\rm dc}+\varphi_d(t)\), with \(\Phi_0\) being the flux quantum~\cite{blais2021circuit} and $\varphi_{\rm dc}$ being the dc bias. We set $\varphi_{\rm dc} = \pi$ for symmetric SQUID. Here, $H_{\rm lin}$ captures the individual contributions from qubits, coupler, charger mediated qubit-coupler and static qubit-qubit interactions:
\begin{equation}
    H_{\rm lin} = H_q + H_{c} + H_{qc} + H_{qq}.
\end{equation}
Each qubit is restricted to its two lowest states, with \(b_i=|0_i\rangle\langle1_i|\), \(X_i=b_i+b_i^\dagger\), and \(Y_i=i(b_i^\dagger-b_i)\) such that the qubit Hamiltonian is
\begin{equation}
    H_q = \sum_{i=1}^4 \frac{\omega_{q,i}}{2\pi} b_i^\dagger b_i.
\end{equation}
The SQUID coupler is described as the oscillator for two modes $\varphi_{\alpha =\pm} = \varphi_a \pm \varphi_b$
\begin{equation}
    H_c = \sum_{\alpha = \pm} \frac{\omega_\alpha}{2\pi} d_\alpha^\dagger d_\alpha
\end{equation}
where the operator \(d_\alpha\) removes one excitation from coupler mode \(\alpha\). The charge-mediated qubit–coupler interaction is
\begin{equation}
\label{eq:qubit-coupler}
H_{qc}=-s_{qc}\sum_{i=1}^{4} Y_i\left(s_i g_{i,-}p_-+g_{i,+}p_+\right)
\end{equation}
where $p_\alpha = i(d_\alpha^\dagger - d_\alpha)$, $g_{i,-} = \sqrt{\chi_i|\omega_{q,i} - \omega_-|/(2\pi)}$ and $g_{i,+} = g_{i,-} \sqrt{\omega_-/\omega_+}$ with $\chi_i$ being the dispersive shift attributed to off-resonant coupling. Here, $s_{qc}$ describes the fraction of charge-coupler coupling strength retained in the calculation. Additionally, $s_i=\{1,-1,1,-1\}$ are relative coupling signs such that qubits connected to opposite terminals contribute with opposite signs. The static qubit-qubit coupling comes from inductive energy of the circuit and can be written as 
\begin{equation}
    H_{qq} = - s_{qq} \frac{\omega_-/(2\pi)}{2\varphi_{-}^2} \left(\varphi_1\varphi_3 X_1 X_3 + \varphi_2\varphi_4 X_2 X_4 \right)
\end{equation}
where $\varphi_{\rm zpf}$ is zero point fluctuation flux amplitude corresponding to mode with energy $\omega_-$ and $\varphi_i$ is flux amplitude through each qubit. As in Eq.~\ref{eq:qubit-coupler}, $s_{qq}$ is the fraction of qubit-qubit static coupling strength retained in the calculation.

Importantly, four-qubit interactions appear from the nonlinearities in the Josephson term of $H_{\rm cell}$ (second term in Eq.~\ref{eq:cell}) that can extracted by expanding $\cos\widehat{\varphi}$. However, for our simulation we retain the full Hamiltonian with quantized flux
\begin{equation}
    \widehat{\varphi} = \varphi_{\rm zpf}(d_- + d_-^\dagger) + \frac{1}{2}\sum_{i=1}^4s_i\varphi_iX_i
\end{equation}
and drive
\begin{equation}
 \varphi_d(t)=\sum_\mu A_\mu
 \cos(\omega_{d,\mu}t+\theta_\mu), \,
 \omega_{d,\mu}=\left|\sum_i\sigma_{\mu,i}\omega_{q,i}\right|
 \label{eq:pump}
\end{equation}
where \(\sigma_{\mu i}=\pm1\) distinguishes excitation from de-excitation, and \(\theta_\mu\) is the tone phase. In rotating frame, the oscillating phase of the selected transition and its reverse, giving stationary interaction terms. The parameters corresponding to Hamiltonian in Eq.~\ref{eq:cell} used in our simulations are shown in \cref{tab:provenance}.

To see how dynamical evolution of Eq.~\ref{eq:cell} generates the required four-qubit interaction, we consider the target interaction $X_1,X_2,X_3,X_4$. Such a four-qubit term contains \(2^4=16\) transitions. Writing \(\sigma_i=+1\) for excitation and \(\sigma_i=-1\) for de-excitation, the net energy change is
\(\hbar\Omega_{\boldsymbol{\sigma}}\). In the interaction picture this gives $\prod_{i=1}^{4} b_i^{(\sigma_i)}(t)
 =e^{i(\sum_{i}\sigma_i\omega_{q,i})t}  \prod_{i=1}^{4} b_i^{(\sigma_i)} $. Since producing \(XXXX\) must flip all four qubits, implementing it requires the complete set of sixteen transitions. The tone amplitudes and phases are chosen so that their stationary coefficients equal the requested value \(-c\):
\begin{equation}
 -c\sum_{\boldsymbol{\sigma}\in\{\pm1\}^{4}}
 \prod_{i=1}^{4} b_i^{(\sigma_i)}
 =-cX_1X_2X_3X_4,
 \label{eq:resonant-stabilizer}
\end{equation}
and choosing other relative tone phases can help realize other four-qubit interactions. 

For numerical simulation, we use the circuit topology, four qubit frequencies, lower SQUID-mode parameters, Josephson energy, and dispersive-coupling scales as in Ref.~\cite{Sameti2017}. We represent the circuit using two states per qubit, 12 lower-mode Fock states, and two upper-mode Fock states. An agent then orchestrates an exhaustive search over 144 predefined circuit settings (shown in Table~\ref{tab:provenance}) using a fixed workflow drawn from recent analyses of driven superconducting circuits~\cite{Rajabzadeh2023,Xiao2025,Dai2026,Lu2026}. For each of the 144 candidate device configurations in the declared search, the agent executes a deterministic one-period simulation of the truncated SQUID circuit~\cite{Petiziol2021,Xiao2025,Dai2026}. The system involves from state $|\psi_\alpha\rangle\rightarrow |\psi_\beta\rangle$ with matrix element $M_{\beta\alpha}$. Here, $\alpha,\beta\in 1,2,...,16$ such that $M$ forms a $16\times 16$ matrix. The transitions outside computational space defined by $\alpha,\beta$ constitute leakage, and we define the maximum escape probability as $ p_{\rm esc}=1-\sigma_{\min}(M)^2$ where $\sigma_{\min}(M)$ is the smallest singular value of $M$~\cite{wood2018quantification,paolo2019control,teske2022qopt} (also see the SM \cref{sm:sec-circuit} for details). The nearest unitary part of \(M\) is finally converted into a complete effective Pauli Hamiltonian. 

The agent then passes its rationalized coefficients and norm witnesses to Lean, which verifies the exact array assembly, bounds the full residual in Eq.~\ref{eq:normal-form}, and proves whether the encoded-sector criterion is satisfied. In other words, the compiler requests that each cell reproduces one stabilizer term of
\(H_0\) with the common interaction strength \(c\), then the circuit calculation
returns 1) the delivered strength \(h_{\rm tar}\) of four-body interaction, 2) $p_{\rm esc}$, and 3) every retained
parasitic Pauli interaction for the chosen pump amplitude, 4) the strengths of the unwanted couplings retained in the model, and 5) the relative phases with which the four qubits couple to the SQUID, as specified in Table~\ref{tab:provenance}.
The agent assembles these complete cell
outputs on the shared qubits, and Lean verifies whether the resulting
\(\effH\) differs from \(H_0\) by a residual small enough to satisfy \cref{thm:criterion}.

\begin{table}[t]
 \caption{Parameters defining the circuits search for single cell configuration. A literature input is transcribed from Ref.~\cite{Sameti2017}. A model completion or selected grid point is an explicit assumption of this demonstration. Further details of the circuit model and numerical procedure are given in SM \cref{sm:sec-circuit} and \cref{sm:sec-cell-search}.
 }
 \label{tab:provenance}
 \scriptsize
{\footnotesize
\setlength{\tabcolsep}{3pt}
\renewcommand{\arraystretch}{1.18}
\begin{ruledtabular}
\begin{tabular}{
  @{}p{0.4\columnwidth}
  p{0.64\columnwidth}@{}
}
Quantity & Value\\
\hline

$\omega_{q,i}/2\pi$
&
$[13.8,13.0,3.5,4.1]\,\mathrm{GHz}$~\cite{Sameti2017}
\\

$\omega_-/2\pi,\ \omega_+/2\pi$
&
$8.45\,\mathrm{GHz}$, $40.0\,\mathrm{GHz}$~\cite{Sameti2017}
\\

$\varphi_{\rm zpf}$
&
$1.65$~\cite{Sameti2017}
\\

$E_J/h$ (Josephson energy)\newline
$\chi_i$ (dispersive shift)
&
$E_J/h=10.0\,\mathrm{GHz}$~\cite{Sameti2017};\newline
$\chi_i=[0.05,0.05,0.025,0.025]\,\mathrm{GHz}$~\cite{Sameti2017}
\\

Pump frequencies
$\omega_{d,\mu}/2\pi$
&
$[0.2,1.4,6.8,8.4, 19.2,26.2,27.4,34.4]\,\mathrm{GHz}$
\\

Selected single-cell design
&
Pump phase: $A_d=0.00470$ per tone;\newline $s_{qc} =2\%$;\newline $s_{qq} = 20\%$;\newline
$\varphi_i=0.7$ for all four qubits
\\

Fine Hilbert truncation
&
$2^4\times12\times2=384$ states
\\

Period and timestep
&
$T=10.0\,\mathrm{ns}$;\newline
$\delta t=0.005\,\mathrm{ns}$ (screening),\newline $\delta t=0.00125\,\mathrm{ns}$ (selected compute)
\\

\end{tabular}
\end{ruledtabular}
}
\end{table}

The $XXXX$ interaction is demanded with strength to be above \(80\,\mathrm{kHz}\), while keeping the largest unwanted driven and static interaction magnitudes to be below \(20\%\) and \(1\%\) of the target, respectively, and \(p_{\rm esc} < 1\%\), more details can be found in the SM \cref{sm:sec-cell-search}.
The requested \(80\,\mathrm{kHz}\) interaction lies below the
MHz-scale four-body couplings proposed for this
architecture~\cite{Sameti2017}. At this target strength, our \(1\%\) static-interaction budget corresponds to \(0.8\,\mathrm{kHz}\). Experiments in other superconducting architectures demonstrate that residual \(ZZ\) coupling can be suppressed into the sub-kHz regime~\cite{Kumph2024RIP,Zhang2024FluxoniumCoupler}. These provide physical scale benchmarks, and the \(20\%\) driven-interaction limit and \(p_{\rm esc} = 1\%\) are specified design budgets. To make the exhaustive search computationally feasible, we adopted the coarse-screening and refinement strategy commonly used in automated superconducting-circuit design~\cite{Menke2021,SQcircuit2023,shanto2024squadds}. Specifically, all 144 settings were first evaluated with eight lower-mode states and a \(0.005\,\mathrm{ns}\) timestep using coarse-to-fine margins calibrated from converged runs. A chosen circuit from the surviving design classes (see \cref{tab:provenance}) was then recomputed with 12 states and a \(0.00125\,\mathrm{ns}\) timestep. The chosen circuit delivered \(h_{\rm tar}=80.916\,\mathrm{kHz}\), with driven and static unwated correlations and $p_{\rm esc}$ of \(11.4\%\), \(0.641\%\), and \(0.974\%\), respectively. This single cell demonstration establishes that a useful four-body scale is locally attainable and we extend this below to periodic toric-code arrays.

\begin{figure*}
    \centering
    \includegraphics[width=\linewidth]{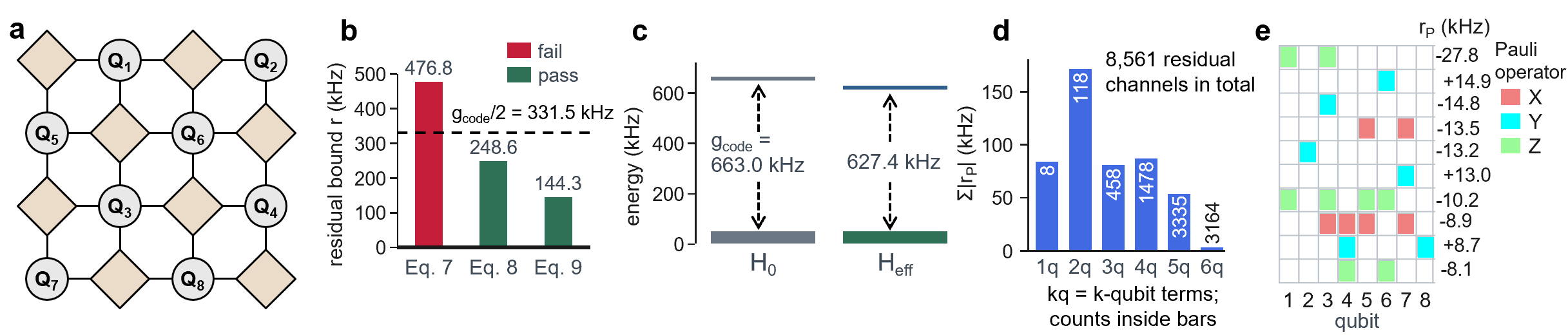}
    \caption{Map of $L=2$ toric code on a SQUID coupled eight qubits. (a) Periodic qubit layout for $L=2$ toric code where brown diamonds represent the SQUID device. (b) Theorem~\ref{thm:criterion} tested on $H_{\rm eff}$ of the selected circuit with criterion described in Eq.~\ref{eq:coefficientwise}-\ref{eq:fullcheck}. It establishes that retaining the interaction algebra [Eq.~\ref{eq:anticommuting}, \cref{thm:anticommuting-bound}] gives a bound closer to the dense witness [Eq.~\ref{eq:fullcheck}]. For circuit design, this implies that grouping of anticommuting Pauli residual interactions in \cref{thm:anticommuting-bound} can help operate at device settings which avoid over compensation solely by adding strength of residual Pauli interactions. (c) Comparison of the toric-code spectrum from $H_0$ and $H_{\rm eff}$ of the selected circuit. Four states remain isolated and close to the intended code space. (d) Aggregate resolution of the full 8561-residual Pauli channels into one-, two-, three-, four-, five- and six-qubit interactions central to evaluating bounds in Eq.~\ref{eq:coefficientwise} and Eq.~\ref{thm:anticommuting-bound} without diagonilizing the full $R$. Bar heights give the total coupling magnitude $\sum |r_a|$, where $r_a$ satisfies $R=\sum_a r_aP_a$ (also see \cref{sec:criterion}). The numbers inside the bars give channel counts. Summing the bars gives the coefficient-wise bound of Eq.~\ref{eq:coefficientwise}. The same coefficients, grouped by anticommutation, enter \cref{thm:anticommuting-bound} and Eq.~\ref{eq:anticommuting}. (e) The ten individual residual Pauli channels with largest $|r_a|$, identifying the dominant residual couplings in the selected circuit. Such a breakdown can be utilized for optimizing coupler designs to suppress residual channels.}
    \label{fig2}
\end{figure*}

\subsection{$L=2$ toric code}
The single-cell experiment does not by itself generate the toric-code space.
For the array-level test, we consider the periodic \(L=2\) torus, where four
star and four plaquette couplers share eight physical edge qubits
\cite{Kitaev2003,Dennis2002,Sameti2017}. The corresponding qubit layout is shown in \cref{fig2}a.  Because neighboring couplers act on common qubits, their responses cannot be assessed independently.  Following the effective-Hamiltonian treatment of driven superconducting circuits
\cite{SQcircuit2023,Xiao2025}, the circuit solver 
calculated each coupler's driven evolutions and extracts the resulting interactions, including qubit-frequency shifts, pairwise couplings, higher-body
parasitics, and the requested four-qubit check.  These interactions are then
placed on the eight physical qubits, where coefficients describing the same
operator are combined, increasing or decreasing its contribution.  We additionally simulate
all 24 overlapping pairs of couplers operating simultaneously and subtract
the responses from each cell operating separately.  The remainder is the connected two-cell correction: effective interactions caused by operating the couplers together~\cite{MuhlhauserSchmidt2022}. The model therefore captures the leading interaction generated specifically by 
nearest-neighboring couplers together.  Three-cell and higher connected processes and cross-cell virtual SQUID-mode excitation remain outside the present calculation as they are expected to be suppressed in off-resonant regime~\cite{Sameti2017,YangBuecheleTrivedi2025}.

We sampled 165 circuit settings by varying the device's capacitance, inductance, Josephson energy, drive amplitude, upper-mode
frequency, retained momentum, and static-pair couplings, as well as the participation-phase
equalization of its qubits. Humans specified which circuit parameters could vary and their permitted ranges. Within these limits, the agent then used seeded Latin-hypercube sampling to distribute initial trials across each parameter range, and then sampled more densely around the settings that best satisfied the numerical acceptance conditions. The scanned parameters are listed in the SM \cref{sm:sec-array} and \cref{sm:tab-array-domain}. 
We performed a numerical test using four lower-mode oscillator states, two upper-mode oscillator states, and a time step of \(0.01\,\mathrm{ns}\). This test shortlisted 11 candidates, which were ranked by the preliminary Hamiltonian conditions listed in \cref{sm:tab-array-testing-conditions}. The selected candidates were recomputed using eight lower-mode states and a \(0.0025\,\mathrm{ns}\) timestep. At finer resolution, two of these configurations satisfied both the leakage requirement ($p_{\rm esc}=1\%$) and \cref{thm:criterion} numerically. 

 Out of the two candidates that satisfy \(2\normop{R}<g_{\rm code}\), we select the one with the larger calculated value of \(g_{\rm code}-2\normop{R}\) 
 (see \cref{sm:tab-array-domain} for the values of selected configuration). Simulating Eq.~\ref{eq:cell} separately for the four star and four plaquette couplers gives extracted strengths ranging from \(165.742\) to \(175.942\,\mathrm{kHz}\) or their requested four-qubit interactions, before connected two-cell corrections are included. To compare these unequal couplings with the uniform toric-code target in Eq.~\ref{eq:target}, we assign every target check the weakest isolated-cell strength of \(c=165.742\,\mathrm{kHz}\) and include all differences and connected two-cell corrections in \(R\). Eq.~\ref{eq:ideal-gap} then gives the ideal target gap \(g_{\rm code}=662.966\,\mathrm{kHz}\). 
 The target coupling strength \(c\) can be a user specified parameter before searching for a suitable circuit. For the current demonstration of the workflow, we instead determine \(c\) from the simulated device response, rather than assuming a value in advance.
 After adding errors due to numerical uncertainty and finite Hilbert space truncation (see Eq.~\ref{eq:criterion}), we find that the coefficient-wise bound in Eq.~\ref{eq:coefficientwise} does not satisfy Eq.~\ref{eq:criterion} and \cref{thm:criterion}. This does not show that the circuit fails, however, adding individual interaction strengths of residual interactions overestimates their possible combined effect. On the contrary, grouping anticommunitng residual interactions, as in \cref{thm:anticommuting-bound} and Eq.~\ref{eq:anticommuting}, gives a smaller upper bound on $\normop{R}$, now satisfying \(r<g_{\rm code}/2\), and therefore establishes the isolated low-energy sector required by Theorem~\ref{thm:criterion} (Fig.~\ref{fig2}b).
 We note that, as each fixed Hamiltonian is evaluated deterministically, whether a residual bound satisfies Theorem~\ref{thm:criterion} is likewise a deterministic calculation outcome. Fig.~\ref{fig2}b implies that during circuit design, grouping of anticommuting residual interactions can retain coupler and pump settings that already produce the requested isolated sector, avoiding additional compensation or retuning demanded solely by the adding strength of residual interactions. Both the numerically obtained spectrum and the spectrum obtained by subtracting the residual are shown in Fig.~\ref{fig2}c. As shown in Fig.~\ref{fig2}b, we further evaluate the exact witness by diagonalizing $R$ (see Eq.~\ref{eq:fullcheck}) which validates that grouping anticommuting residual interactions imposes a tighter bound than simply adding the individual strength of residual interactions. 

Moreover, the projector estimate in Eq.~\ref{eq:projector-bound} bounds
\(\normop{P_{\rm low}-UP_0U^\dagger}\) which, by
diagonalization of the central Hamiltonian, evaluates to \(0.054\). For $r/(g_{\rm code}-r)$, we use estimates from the witness evaluating to 0.279. Thus, Eq.~\ref{eq:projector-bound} in \cref{thm:criterion} is satisfied in our demonstration. This implies that the isolated states remain close to the intended toric-code space, with the numerical result showing less mixing with states outside that space than the theorem's upper bound permits.

The complete residual anatomy is shown in Fig.~\ref{fig2}.d and Fig.~\ref{fig2}.e.  Two-qubit channels carry the largest aggregate coefficient mass, \(170.750\,\mathrm{kHz}\). For the selected configuration, the largest unwanted interaction is a two-qubit ZZ coupling of 27.841 kHz, approximately 17\% of the requested stabilizer strength, identifying a concrete target for further coupler optimization.
Several one-qubit \(Y\) fields are also among the leading channels, while the connected two-cell calculation generates smaller three- through six-qubit terms that are absent from an isolated-cell description. These couplings are calculated deterministically for the selected circuit and identify its leading targets for improvement, but their magnitudes and relative importance can change with the circuit parameters. This converts the Lean checked workflow into a hardware diagnostic: it identifies which couplings must be suppressed, while the syndrome and logical class of each Pauli word
provide the input for subsequent analyses of detectable excitations and coherent logical evolution.

\section{Conclusion}

In this work we develop an agentic pipeline that certifies whether a declared circuit Hamiltonian supports an isolated low-energy sector with the required number of states and a bounded deviation from the intended topological-code space.
The pipeline derives and formalizes the theorems that are sufficient conditions for an isolated low-energy sector on the device. We demonstrate the validity of our method by implementing a $L=2$ toric code on a device with driven-SQUID couplers. The code specifies an encoded many-body sector, the driven-SQUID model supplies that target together with thousands of coherent parasitic channels, and the residual operator determines whether energy still identifies the requested subspace. We comment that while this work relies on numerically simulated data, the pipeline can be adapted for device simulation and experimental data.

Operationally, 
our method closes several concrete loops between code design and hardware. Before fabrication, a superconducting-circuit optimizer can vary junction energies, mode frequencies, participation factors, pump amplitudes, and compensation channels 
to increase the energy separation between the desired low-energy states and the remaining states, rather than only strengthening the desired four-qubit interaction \cite{Menke2021,Eriksson2026}.
An analog compiler can similarly translate a surface-code or color-code Hamiltonian into native couplers and drives, then submit the complete compiled Hamiltonian for verification \cite{SimuQ2024,VerifiedCompiler2025}. Importantly, grouping of anticommuting Pauli interactions ( \cref{thm:anticommuting-bound}), applied for the first time to QEC, allows the residual bound to be evaluated from the Pauli inventory without diagonalizing the full array Hamiltonian. After fabrication, Hamiltonian-learning or spectroscopy data can replace simulated coefficients, turning the same proof into a device-level acceptance test and identifying spatially localized implementation errors \cite{Hangleiter2024}. Our framework could compare code layouts and establish operating ranges across bounded fabrication variations, connecting device design to experimentally assessable requirements.

Beyond finite-system realization, the decomposed residual interactions could guide hardware optimization by distinguishing interactions that create excitations from those that take the system out of the code space~\cite{BravyiHastingsMichalakis2010,LiODeaKhemani2025}. Coupler redesign and selective control could then mitigate these effects \cite{Choi2020,Nguyen2026}.  The agentic workflow could automate the search for such improvements, while Lean checks whether each updated Hamiltonian satisfies the established spectral conditions, allowing the Lean codes to be reused for successive design changes. Combining the reconstructed Hamiltonian with noise and recovery models would also allow comparisons with and without stabilizer energy penalties. This would test whether these interactions reduce errors or allow the qubits to be checked less frequently~\cite{YoungSarovarBlumeKohout2013,ZhengBrun2015}.

Such extensions would connect the present results to scalable Hamiltonian engineering and the operational benefits of hybrid analog--active QEC. Additionally, the same workflow can be applied to address hardware concerns in digital QEC, e.g., connectivity, latency, and crosstalk. Broadly, the certified Hamiltonian and residual inventory can then feed downstream studies of logical splitting, state preparation, syndrome dynamics, decoder performance, and open-system memory lifetime. Our work lays a foundation for automated workflows where AI \addAA{agents} translate a desired quantum behavior into device designs and refine them through simulation and experiment with reusable underlying mathematical reasoning through Lean proofs. Ultimately, this could help turn quantum hardware development into predictive engineering, where new architectures are inverse designed for the many-body behavior they can reliably deliver, accelerating the translation of theoretical advances into experimentally validated quantum technologies.

\section*{acknowledgments}
As detailed in \cref{sec:agentic-workflow}, OpenAI Codex was used to develop mathematical arguments, generate Lean and numerical code, execute simulations and parameter searches, prepare figures, and draft and revise manuscript text. Anthropic Claude provided additional critical review of the scientific arguments and presentation. Large language models were also used for writing assistance. The human authors directed the research, reviewed and revised the outputs, and retain responsibility for the scientific claims and final manuscript. \\

\section*{Data and code availability}
The Lean proof files, supporting numerical records, and verification code will be made publicly available upon publication of this article.

\bibliography{references}


\ifdefined\SMStandaloneBuild
\else
  \clearpage
  \onecolumngrid
  \begin{center}
    {\large\bfseries Supplemental Material}
  \end{center}
  \vspace{0.5em}
\fi

\setcounter{section}{0}
\setcounter{subsection}{0}
\setcounter{equation}{0}
\setcounter{figure}{0}
\setcounter{table}{0}
\renewcommand{\thesection}{S\arabic{section}}
\renewcommand{\thesubsection}{\thesection.\arabic{subsection}}
\renewcommand{\theequation}{S\arabic{equation}}
\renewcommand{\thefigure}{S\arabic{figure}}
\renewcommand{\thetable}{S\arabic{table}}

\renewcommand{\theHequation}{SM.\arabic{equation}}
\renewcommand{\theHtable}{SM.\arabic{table}}

This Supplemental Material gives the mathematical proofs, circuit-reduction
procedure, search contract, numerical ledger, and formal-verification boundary
supporting the main text. Frequencies are ordinary frequencies rather than
angular frequencies unless an explicit factor of $2\pi$ is shown. Numerical
Hamiltonian coefficients are stored in GHz and converted to kHz for
presentation. The circuit calculation proposes coefficients and witnesses;
Lean checks exact rational consequences of those imported objects. It does
not prove that the declared model completely describes a fabricated device.

\section*{Supplemental contents}
\begingroup
\setlength{\parindent}{0pt}
\newcommand{\SMcontentsline}[2]{\noindent\hyperref[#1]{#2}\dotfill\pageref{#1}\par}
\SMcontentsline{sm:sec-code}{S1\quad Code geometry, Hamiltonian, and ideal gap}
\SMcontentsline{sm:sec-criterion}{S2\quad Analytical proof of Theorem 1 in the main text}
\SMcontentsline{sm:sec-bounds}{S3\quad Derivation of the residual-norm certificates}
\SMcontentsline{sm:sec-circuit}{S4\quad From driven-circuit dynamics to an operational Hamiltonian}
\SMcontentsline{sm:sec-cell-search}{S5\quad Single-cell search and refinement}
\SMcontentsline{sm:sec-array}{S6\quad Eight-qubit search and connected two-cell model}
\SMcontentsline{sm:sec-anatomy}{S7\quad Residual anatomy and microscopic provenance}
\SMcontentsline{sm:sec-guard}{S8\quad Numerical guard and model boundary}
\SMcontentsline{sm:sec-result}{S9\quad Final exact certificate and central diagnostics}
\SMcontentsline{sm:sec-formal}{S10\quad Formalization, agentic workflow, and trust boundary}
\endgroup

\section{Code geometry, Hamiltonian, and ideal gap}
\label{sm:sec-code}

Let $C_2\xrightarrow{\partial_2}C_1\xrightarrow{\partial_1}C_0$ be the
cellular chain complex of an $L\times L$ periodic square lattice. Faces,
edges, and vertices generate $C_2$, $C_1$, and $C_0$, respectively, and
$\partial_1\partial_2=0$. Reduction modulo two gives binary incidence maps
$B_2$ and $B_1$. One qubit is placed on each edge. For every vertex $v$ and
face $f$, define
\begin{equation}
 A_v=\prod_{e\in E}X_e^{B_1(v,e)},
 \qquad B_f=\prod_{e\in E}Z_e^{B_2(e,f)}.
 \label{sm:eq-checks}
\end{equation}
The chain condition implies $B_1B_2=0$, so every star and plaquette overlap on
an even number of qubits and commute. The positive-projector Hamiltonian is
\begin{equation}
 \codeH=J_X\sum_v\frac{\id-A_v}{2}
 +J_Z\sum_f\frac{\id-B_f}{2}.
 \label{sm:eq-projector-hamiltonian}
\end{equation}
Its kernel is precisely the simultaneous $+1$ eigenspace of the checks. In
the scalar-free compiler convention, $J_X=J_Z=2c$ and
\begin{equation}
 H_0(c)=-c\left(\sum_v A_v+\sum_f B_f\right).
 \label{sm:eq-target}
\end{equation}
The two forms differ only by a scalar and have identical eigenvectors and
gaps.

On a torus, the product of all stars and the product of all plaquettes are
identities. The $2L^2$ checks therefore have two global relations, while two
independent noncontractible cycles label two encoded qubits. Thus the ground
sector has dimension $K=4$. Flipping one check eigenvalue is impossible on a
closed periodic lattice: electric and magnetic defects are created in pairs.
The smallest excited sector of Eq.~\eqref{sm:eq-target} changes two check
eigenvalues from $+1$ to $-1$. Each change costs $2c$, giving
\begin{equation}
 g_{\rm code}=4c.
 \label{sm:eq-gap}
\end{equation}
For a nonuniform delivered array we still request one uniform $H_0(c)$: $c$
is the smallest delivered check magnitude, and every excess or deficit of an
individual check coefficient remains in the residual. This preserves
Eq.~\eqref{sm:eq-gap} rather than replacing the target by a weighted model.

For $L=2$, the qubit order is \((h_{00},h_{01},v_{00},v_{01},h_{10},h_{11},v_{10},v_{11})\), equivalent to
labels $1,\ldots,8$. The check supports are listed below.
\begin{table}[t]
\caption{Eight check supports in the periodic $L=2$ array. The source driving
convention produces $XXXX$ star cells and $YYYY$ plaquette cells; the declared
local frame maps the latter to the CSS $Z$ convention.}
\label{sm:tab-supports}
\begin{ruledtabular}
\begin{tabular}{lcl@{\qquad}lcl}
Cell & Type & Qubits & Cell & Type & Qubits \\
\hline
star-00 & $X$ & $1,2,3,7$ & plaquette-00 & $Z$ & $1,3,4,5$ \\
star-01 & $X$ & $1,2,4,8$ & plaquette-01 & $Z$ & $2,3,4,6$ \\
star-10 & $X$ & $3,5,6,7$ & plaquette-10 & $Z$ & $1,5,7,8$ \\
star-11 & $X$ & $4,5,6,8$ & plaquette-11 & $Z$ & $2,6,7,8$ \\
\end{tabular}
\end{ruledtabular}
\end{table}

\section{Analytical proof of Theorem 1 in the main text}
\label{sm:sec-criterion}

Theorem 1 is not a new perturbation theorem. It is a finite-dimensional
specialization of standard eigenvalue and spectral-subspace perturbation
theory \cite{Kato1995,Bhatia1997,DavisKahan1970}. Gap stability is already
central to Hamiltonian formulations of topological QEC
\cite{Bravyi2010}, while control of the implemented code subspace is the
starting point for approximate-QEC statements
\cite{KnillLaflamme1997,BenyOreshkov2010}. The contribution of the main text
is therefore not the underlying matrix inequality, but its use as an exact,
machine-checked interface between a prescribed code Hamiltonian and a complete
declared circuit Hamiltonian. We include the derivation because this interface
is the logical step on which the hardware certificate depends.

Let $H_0$ be Hermitian and positive semidefinite, let its ground projector
$P_0$ have rank $K$, and suppose its remaining spectrum begins at
$g_{\rm code}>0$. Let $U$ be an allowed product of local basis changes and
write
\begin{equation}
 \effH=UH_0U^\dagger+R,\qquad R=R^\dagger,\qquad \normop{R}\le r.
 \label{sm:eq-normal-form}
\end{equation}
Theorem 1 asserts that, if $2r<g_{\rm code}$, exactly $K$ eigenvalues of
$\effH$ lie below $g_{\rm code}/2$, no eigenvalue lies in
$(r,g_{\rm code}-r)$, and the separation between the low and high groups is
at least $g_{\rm code}-2r$. If $P_{\rm low}$ projects onto the low group, it
also asserts
\begin{equation}
 \normop{P_{\rm low}-UP_0U^\dagger}\le\frac{r}{g_{\rm code}-r}.
 \label{sm:eq-projector}
\end{equation}

\subsection{Counting the low-energy states}

Set $A=UH_0U^\dagger$, $P=UP_0U^\dagger$, and $B=\effH=A+R$. For every
normalized state $\lvert\psi\rangle$,
$|\langle\psi|R|\psi\rangle|\le r$. The variational characterization of
ordered eigenvalues therefore says that adding $R$ can move any eigenvalue of
$A$ upward or downward by at most $r$. Since the first $K$ eigenvalues of $A$
are zero and every remaining eigenvalue is at least $g_{\rm code}$, one has
\begin{equation}
 \lambda_i(B)\le r\ (i\le K),\qquad
 \lambda_i(B)\ge g_{\rm code}-r\ (i>K).
 \label{sm:eq-weyl}
\end{equation}
Since $r<g_{\rm code}/2<g_{\rm code}-r$, exactly $K$ eigenvalues lie below
the midpoint. The interval $(r,g_{\rm code}-r)$ is empty, and the distance
between the highest member of the low group and the lowest member of the high
group is at least $g_{\rm code}-2r$. This proves the state-count and spectral
separation parts of Theorem 1 without assuming that the perturbation preserves
any stabilizer or symmetry.

\subsection{Bounding the rotation of the code space}

Let $Q=P_{\rm low}$, let $A_\perp$ restrict $A$ to $\im(P)^\perp$, and let
$B_Q$ restrict $B$ to $\im(Q)$. The operator
$X=(\id-P)Q$ is the component of the physical low-energy space that points
outside the ideal code space. Because $Q$ is a spectral projector of $B$, it
commutes with $B$. Projecting $B=A+R$ from the left with $\id-P$ and from the
right with $Q$ gives
\begin{equation}
 A_\perp X-XB_Q=-(\id-P)RQ.
 \label{sm:eq-sylvester}
\end{equation}
The operators on the two sides of $X$ act in separated energy windows:
$A_\perp\ge g_{\rm code}\id$, whereas every eigenvalue of $B_Q$ is at most
$r$. Equation~\eqref{sm:eq-sylvester} can consequently be solved directly as
\begin{equation}
 X=-\int_0^\infty e^{-tA_\perp}(\id-P)RQe^{tB_Q}\,dt .
 \label{sm:eq-projector-integral}
\end{equation}
Differentiating the integrand verifies Eq.~\eqref{sm:eq-sylvester}; the upper
limit vanishes because $g_{\rm code}>r$. Taking norms inside the integral and
using $\normop{e^{-tA_\perp}}\le e^{-tg_{\rm code}}$,
$\normop{e^{tB_Q}}\le e^{tr}$, and
$\normop{(\id-P)RQ}\le r$ yields
\begin{equation}
 \normop{X}
 \le r\int_0^\infty e^{-t(g_{\rm code}-r)}dt
 =\frac{r}{g_{\rm code}-r}.
 \label{sm:eq-sylvester-bound}
\end{equation}
The first part of the proof established $\operatorname{rank}P=
\operatorname{rank}Q=K$. For equal-rank orthogonal projectors, their operator
distance is the sine of the largest principal angle and equals
$\normop{(\id-P)Q}=\normop X$. Equation~\eqref{sm:eq-sylvester-bound} is
therefore exactly Eq.~\eqref{sm:eq-projector}. Physically, it limits how far
the entire implemented low-energy subspace can rotate away from the requested
code space, rather than comparing selected eigenvectors one at a time.

The condition $2r<g_{\rm code}$ is deliberately sufficient rather than
necessary. If it fails, these worst-case intervals overlap and this proof
stops deciding the question; the physical Hamiltonian may still possess an
encoded sector, but establishing it then requires a sharper residual bound or
a different, more structure-sensitive argument. The finite global estimate
used here should also be distinguished from thermodynamic stability theorems
for local topological Hamiltonians \cite{Bravyi2010}, which address a stronger
large-system question under additional locality assumptions.

\section{Derivation of the residual-norm certificates}
\label{sm:sec-bounds}

After fixing the target coefficient and local frame, the complete coherent
mismatch is expanded in the Hermitian $n$-qubit Pauli basis,
\begin{equation}
 R=\sum_{P\in\mathcal P_n}r_PP,
 \qquad \mathcal P_n=\{I,X,Y,Z\}^{\otimes n},
 \label{sm:eq-pauli-residual}
\end{equation}
Here $r_P$ is one signed Pauli coefficient; it is not the scalar $r$ in
main-text Theorem 1, which bounds the norm of the complete sum.
Every Pauli word is Hermitian and unitary, so $P^2=\id$ and
$\normop{P}=1$. Applying the triangle inequality term by term gives the
coefficientwise baseline
\begin{equation}
 \normop{R}
 \le \sum_P\normop{r_PP}
 =\sum_P|r_P|.
 \label{sm:eq-l1}
\end{equation}
This bound is rigorous, but it discards how the channels act relative to one
another: it assumes that one state can simultaneously attain the largest
signed expectation value of every Pauli word. Anticommuting observables cannot
do so, and retaining that fact produces a stronger certificate.

\subsection{Pauli-commutation bound}

A Pauli word can be encoded by two binary strings $x,z\in\Ftwo^n$, where
$x_k=1$ marks an $X$ component and $z_k=1$ marks a $Z$ component on qubit
$k$; $x_k=z_k=1$ represents $Y$. Two words $P(x,z)$ and $P(x',z')$
anticommute exactly when
\begin{equation}
 x\mathbin{\cdot}z'+z\mathbin{\cdot}x'=1\pmod 2.
 \label{sm:eq-symplectic-anticommutation}
\end{equation}
Equation~\eqref{sm:eq-symplectic-anticommutation} is obtained by moving the
single-qubit $X$ and $Z$ factors of one word past those of the other: every
$XZ=-ZX$ encounter contributes one minus sign, and only the parity of the
number of such encounters matters.

Partition the nonzero terms of $R$ into disjoint groups $\mathcal G_j$ such
that every pair of distinct words in the same group anticommutes. Define
$R_j=\sum_{a\in\mathcal G_j}r_aP_a$. Expanding its square gives
\begin{align}
 R_j^2
 &=\sum_{a\in\mathcal G_j}r_a^2P_a^2
 +\sum_{\substack{a,b\in\mathcal G_j\\a<b}}
 r_ar_b(P_aP_b+P_bP_a) \nonumber\\
 &=\left(\sum_{a\in\mathcal G_j}r_a^2\right)\id .
 \label{sm:eq-group-square}
\end{align}
The first equality separates the diagonal and cross terms. In the second,
$P_a^2=\id$ evaluates every diagonal term, while pairwise anticommutation
makes every cross term vanish. Since $R_j$ is Hermitian,
$\normop{R_j}^2=\normop{R_j^2}$, and hence
\begin{equation}
 \normop{R_j}=\sqrt{\sum_{a\in\mathcal G_j}r_a^2}.
 \label{sm:eq-group-norm}
\end{equation}
The triangle inequality is now used only between groups rather than between
individual channels:
\begin{equation}
 \normop R
 =\normop{\sum_jR_j}
 \le\sum_j\normop{R_j}
 =\sum_j\sqrt{\sum_{a\in\mathcal G_j}r_a^2}.
 \label{sm:eq-anticommuting}
\end{equation}
Singleton groups reproduce Eq.~\eqref{sm:eq-l1}; any nontrivial
anticommuting group can only improve it because the Euclidean norm of its
coefficient vector is no larger than its one-norm. In particular, $N$ terms
of equal magnitude $\epsilon$ contribute $\epsilon\sqrt N$ when they
anticommute pairwise, rather than the $N\epsilon$ assigned by the
coefficientwise bound. Anticommuting partitions have previously been used to
reorganize Pauli operators in variational quantum algorithms
\cite{Izmaylov2020,Ralli2021}; here the same exact algebra bounds a coherent
hardware residual.

The partition need not be optimal for the result to be valid. A numerical
routine searches for large anticommuting groups, but the imported certificate
contains the complete list of groups. Lean independently checks that every
nonzero Pauli term occurs exactly once and evaluates
Eq.~\eqref{sm:eq-symplectic-anticommutation} for every pair placed in the same
group. The optimization routine can therefore affect the tightness of the
bound, but not its correctness.

\subsection{Exact dense-matrix witness}

For a small system, the same residual can be checked without choosing a Pauli
partition. Because $R$ is Hermitian, proving $\normop R\le r_{\rm cert}$ is equivalent
to proving that every eigenvalue $\lambda$ of $R$ satisfies
$\lambda^2\le r_{\rm cert}^2$, or, equivalently,
\begin{equation}
 r_{\rm cert}^2\id-R^2\succeq0.
 \label{sm:eq-positive-target}
\end{equation}
A floating-point eigensolver or Cholesky factorization can suggest a useful radius $r_{\rm cert}$, but rounding prevents that numerical output from proving
Eq.~\eqref{sm:eq-positive-target}. We instead rationalize the proposed factor
and verify the following exact identity:
\begin{equation}
 r_{\rm cert}^2\id-R^2=LL^\dagger+\delta\id+E,
 \qquad E=E^\dagger,\quad \delta\ge0,\quad \normop E\le\delta.
 \label{sm:eq-dense-witness}
\end{equation}
Here $L$ is the rationalized candidate factor, and $E$ is not a fitted
remainder: it is defined by subtracting $LL^\dagger+\delta\id$ from the exact
matrix $r_{\rm cert}^2\id-R^2$. For any state $|\psi\rangle$,
\begin{align}
 \langle\psi|(r_{\rm cert}^2\id-R^2)|\psi\rangle
 &=\normop{L^\dagger|\psi\rangle}^2
 +\delta\normop{|\psi\rangle}^2
 +\langle\psi|E|\psi\rangle \nonumber\\
 &\ge \normop{L^\dagger|\psi\rangle}^2
 +(\delta-\normop E)\normop{|\psi\rangle}^2
 \ge0.
 \label{sm:eq-witness-positivity}
\end{align}
Thus Eq.~\eqref{sm:eq-dense-witness} proves
Eq.~\eqref{sm:eq-positive-target}, and diagonalizing $R$ conceptually---not
numerically---then gives $|\lambda|\le\rho$ for every eigenvalue and therefore
$\normop R\le r_{\rm cert}$.

All entries of $R$, $L$, $r_{\rm cert}$, and $\delta$ are represented as exact
rationals after a common scaling. Lean recomputes every entry of $E$ and
bounds its operator norm by its maximum absolute row sum. For Hermitian $E$,
\begin{equation}
 \normop E
 \le\max_i\sum_j|E_{ij}|,
 \label{sm:eq-row-bound}
\end{equation}
so checking that every exact row sum is at most $\delta$ supplies the last
inequality in Eq.~\eqref{sm:eq-dense-witness}. Floating point is used only to
discover a compact candidate $L$; neither the numerical factorization nor its
reported eigenvalues belong to the trusted proof.

\subsection{Memory scaling and the scope of scalability}

The two certificates store fundamentally different objects. Let $m$ be the
number of nonzero Pauli channels, let $g$ be the number of anticommuting
groups, and let $b$ bound the number of bits used for an exact coefficient.
A sparse Pauli certificate stores, for each channel, a $2n$-bit symplectic
label, a coefficient, and a group identifier. Its idealized storage is
therefore
\begin{equation}
 M_{\rm Pauli}=O\!\left(m[2n+b+\log_2g]\right)\ \text{bits}.
 \label{sm:eq-pauli-memory}
\end{equation}
No state vector or $2^n\times2^n$ operator is required. Pairwise
anticommutation can be verified directly from the stored bit masks and the
group lists. If these checks are streamed, they require no asymptotically
larger memory, although their worst-case time is
$O(n\sum_j|\mathcal G_j|^2)\subseteq O(nm^2)$. A heuristic may construct and
store the full $m\times m$ anticommutation graph while searching for a good
partition, but that graph is neither part of the proof witness nor required by
the verifier.

The dense route instead uses matrices of linear dimension $d=2^n$. Storing
either $R$ or its factor $L$ requires $d^2=4^n$ exact entries and hence
\begin{equation}
 M_{\rm dense}=O(b4^n)\ \text{bits},
 \label{sm:eq-dense-memory}
\end{equation}
up to a constant number of matrices. Triangular storage changes only the
prefactor. A conventional dense factorization also takes $O(d^3)=O(8^n)$
arithmetic operations. The exact defect $E$ need not be retained in full if
its rows are reconstructed and checked sequentially, but the factor $L$
itself already imposes the exponential memory scaling.

\begin{table}[t]
\caption{Memory comparison for the two residual certificates. The Pauli
advantage is conditional on a sparse retained Hamiltonian, $m\ll4^n$.}
\label{sm:tab-memory-scaling}
\begin{ruledtabular}
\begin{tabular}{p{0.20\textwidth}p{0.35\textwidth}p{0.31\textwidth}}
Certificate & Stored proof data & Memory scaling \\
\hline
Pauli commutation & $m$ symplectic labels, exact coefficients, and group IDs
& $O(m[n+b+\log g])$ \\
Dense witness & $2^n\times2^n$ residual and rationalized factor
& $O(b4^n)$ \\
\end{tabular}
\end{ruledtabular}
\end{table}

For the present $n=8$ calculation, $d=256$ and a dense matrix contains
$65{,}536$ entries; the certified triangular factor has $32{,}896$ nonzero
Gaussian-integer entries. The sparse representation instead stores $m=8{,}561$
Pauli coefficients, each accompanied by a $16$-bit label and one group index.
Both routes are practical at this size. With each additional qubit, however,
the dense matrix storage grows by a factor of four, whereas an existing Pauli
label grows by only two bits.

This is the precise sense in which the Pauli-channel certificate is scalable:
if a bounded-degree device model is retained to fixed interaction or connected
cluster order, then $m$ grows linearly or polynomially with the array size and
Eq.~\eqref{sm:eq-pauli-memory} remains polynomial. An arbitrary $n$-qubit
Hamiltonian can contain all $4^n$ Pauli words; in that worst case the sparse
description is itself exponential and no scalability claim applies. The
certificate therefore removes the exponential Hilbert-space matrix from the
proof path, but it does not make an intrinsically dense Hamiltonian sparse.

Whichever route supplies the central-model enclosure $\rho$, main-text
Theorem 1 receives
\begin{equation}
 r=r_{\rm cert}+\eta,
 \label{sm:eq-rho-eta}
\end{equation}
where $\rho$ is a proved central-model enclosure and $\eta$ is the separately
declared guard in Sec.~\ref{sm:sec-guard}.

\section{From driven-circuit dynamics to an operational Hamiltonian}
\label{sm:sec-circuit}

The reduction in this section is standard numerical circuit analysis rather
than a contribution of this work. Its utility is that it translates a driven,
multilevel circuit into the language in which a compiler request is stated:
leakage from the computational manifold and a complete effective Hamiltonian
acting within it. Variants of this translation underpin automated coupler
design \cite{Menke2021}, general superconducting-circuit analysis
\cite{SQcircuit2023}, Floquet interaction engineering
\cite{Nguyen2024Floquet,Ding2024QICK}, leakage-aware pulse design
\cite{Hyyppa2024}, and processor-scale Hamiltonian reconstruction
\cite{Hangleiter2024Hardware}.

The coupling cell contains four frequency-distinct qubits and two auxiliary
SQUID normal modes. Its time-dependent circuit Hamiltonian is represented
schematically by
\begin{equation}
 H_{\rm cell}(t)=H_{\rm lin}
 -E_J\cos\!\left[\varphi_{\rm dc}+\widehat\varphi+\varphi_d(t)\right],
 \label{sm:eq-cell}
\end{equation}
and each term has a distinct engineering role. The linear part $H_{\rm lin}$
sets the dressed mode frequencies and the static hybridization through which
the qubits participate in the coupler. The Josephson scale $E_J$ fixes the
available nonlinear interaction strength. The dc bias $\varphi_{\rm dc}$ selects
the operating point and controls which orders of the cosine expansion are
enhanced or suppressed. The operator $\widehat\varphi$ contains the zero-point
flux participation of every qubit and auxiliary mode, thereby setting the
matrix element of each mixing process. Finally, the multitone drive
$\varphi_d(t)$ supplies frequency, phase, and amplitude as independent control
knobs: its frequency brings a chosen creation--annihilation process into
resonance, its phase selects the Pauli quadrature, and its amplitude controls
the delivered rate. This is the same resource that enables programmable
parametric exchange and Floquet-engineered spin interactions in contemporary
superconducting devices \cite{Nguyen2024Floquet,Ding2024QICK}.

The important point is that the cosine does not generate only the requested
process. The same expansion produces ac-Stark shifts, residual longitudinal
and transverse two-qubit couplings, and higher-weight interactions. A useful
reduction must therefore retain the whole modeled response rather than read
off one resonant coefficient \cite{Sameti2017}.

Let $V_d$ contain the 16 dressed states assigned to the four-qubit bare basis
with auxiliary modes in vacuum. Projecting the full one-period propagator
$\mathcal U(T)$ onto these columns gives
\begin{equation}
 M=V_d^\dagger\mathcal U(T)V_d,
 \qquad p_{\rm esc}=1-\sigma_{\min}(M)^2.
 \label{sm:eq-projected-map}
\end{equation}
This compression asks the first practical question: does the driven device
remain a four-qubit device? For an input vector $|\psi\rangle$ in the dressed
manifold, $\normop{M|\psi\rangle}^2$ is the probability retained there after
one period. Its minimum over normalized inputs is
$\sigma_{\min}(M)^2$, so $p_{\rm esc}$ is the worst-case loss to auxiliary or
higher circuit levels. Leakage-aware modeling is now essential both for fast
transmon control and for QEC, where leaked population can create correlated
errors not represented by a two-level Hamiltonian \cite{Hyyppa2024}. The
$1\%$ cutoff used below is our declared engineering budget, not a universal
fault-tolerance threshold.

Projection makes $M$ nonunitary whenever population escapes. Writing its
singular-value decomposition as $M=A\Sigma B^\dagger$, the polar unitary
$Q=AB^\dagger$ answers a second question: what coherent transformation remains
after the contraction associated with leakage is separated out? This prevents
population loss from being misreported as a coherent interaction error. We
then remove declared one-body phases that can be compensated by calibrated
virtual-$Z$ frame updates. For the remaining interaction unitary
$Q_{\rm int}$, the principal Floquet logarithm and Pauli projection give
\begin{equation}
 H_F=\frac{i}{2\pi T}\log Q_{\rm int},
 \qquad h_P=2^{-4}\operatorname{Tr}(PH_F).
 \label{sm:eq-floquet}
\end{equation}
The logarithm converts one-period motion into the static generator that would
produce the same stroboscopic evolution. The principal branch is accepted only
when every eigenphase remains away from its cut, so the extracted generator
does not jump by an untracked Floquet zone. Pauli orthogonality then gives each
$h_P$ a direct interpretation: the identity is an irrelevant energy offset;
one-qubit words are local frequency or frame shifts; two-qubit words are
exchange, Ising, or crosstalk channels; and three- and four-qubit words contain
the requested many-body interaction together with coherent parasitic
processes. Static dressed-energy terms are projected into the same basis and
added before array assembly.

Recent experiments illustrate why each output is useful. Floquet drives have
been used to program the anisotropy of effective Heisenberg interactions and
to realize iSWAP, CZ, SWAP, and three-qubit gates
\cite{Nguyen2024Floquet}; parametrically pumped nonlinear couplers activate
selected exchange processes while imposing experimentally relevant phase
constraints \cite{Ding2024QICK}; and learned Pauli Hamiltonians now provide
spatial error maps across superconducting processors
\cite{Hangleiter2024Hardware}. In automated design, the same forward-model
logic turns circuit parameters into objectives that an optimizer can search
\cite{Menke2021,SQcircuit2023}. These works establish the usefulness of the
individual reduction steps; they do not all use the exact sequence above.
Our calculation adopts that established numerical toolkit as its front end.
Its output is the complete Pauli inventory within the declared truncation,
while the new task begins only afterward: certifying whether that inventory
realizes the requested encoded sector.

The agent orchestrates propagation, dressed-state assignment, leakage
evaluation, polar decomposition, branch checking, and Pauli extraction. Lean
receives rationalized coefficients and proof witnesses only after those
standard numerical steps; it certifies their downstream algebraic
consequences, not the adequacy of the circuit model itself.

\section{Single-cell search and refinement}
\label{sm:sec-cell-search}

The main text lists the numerical requirements imposed on the single cell.
Here we explain what the scanned coordinates change physically. Before the
couplings are turned on, a convenient basis is
\begin{equation}
 |q_1q_2q_3q_4;n_-,n_+\rangle,
 \qquad q_i\in\{0,1\},quad n_\pm\in\mathbb N,
 \label{sm:eq-bare-configurations}
\end{equation}
where $q_i$ records the state of qubit $i$ and $n_\pm$ count excitations of
the two SQUID normal modes. Its approximate uncoupled energy is the sum of the
occupied qubit and normal-mode frequencies. Linear hybridization and the
Josephson cosine mix configurations with the same allowed circuit symmetries,
turning crossings of bare levels into avoided crossings and producing dressed
eigenstates
\begin{equation}
 |\widetilde\alpha\rangle
 =\sum_{\bm q,n_-,n_+}c^{(\alpha)}_{\bm q,n_-,n_+}
 |\bm q;n_-,n_+\rangle .
 \label{sm:eq-dressed-configurations}
\end{equation}
The computational manifold consists of the 16 dressed levels continuously
assigned to $|\bm q;0,0\rangle$ for $\bm q\in\{0,1\}^4$. Requiring assignment
overlap above $0.999$ ensures that these levels still behave predominantly as
four-qubit configurations rather than auxiliary-mode excitations.

Periodic driving adds a second level structure. The eigenphases of the
one-period propagator are quasienergies, defined modulo the drive-frequency
grid. A pump at a signed sum of the four qubit frequencies makes one
four-qubit creation--annihilation path stationary in the rotating frame.
Across the 16 computational configurations this appears as the correlated
quasienergy splitting associated with the requested $XXXX$ or $YYYY$ Pauli
word. Off-resonant paths, static pair interactions, and hybridization with
$n_\pm>0$ configurations produce the unwanted Pauli coefficients and the
escape probability extracted in Sec.~\ref{sm:sec-circuit}. Thus the search is
not merely maximizing one transition rate: it is reshaping an entire dressed
and Floquet spectrum while preserving an identifiable four-qubit manifold.

\subsection{Physical role of the search coordinates}

The single-cell experiment asks whether one driven coupler can deliver a
four-body check above $80\khz$ while keeping the largest driven parasitic below
$20\%$ of the target, the modeled static correlated channel below $1\%$,
$p_{\rm esc}<1\%$, and dressed-state overlap above $0.999$. The 144 screening
points are the Cartesian product
\begin{equation}
 \begin{split}
 s_{\rm mom}&\in\{0.02,0.05,0.08\},\quad
 s_{\rm pair}\in\{0,0.1,0.2,0.3\},\\
 e_\phi&\in\{0.90,0.95,1.00\},\quad
 A_d\in\{0.0047,0.0048,0.0049,0.0050\}\ \mathrm{rad}.
 \end{split}
 \label{sm:eq-cell-grid}
\end{equation}

The retained momentum factor $s_{\rm mom}$ multiplies the source-informed
momentum-mediated coupling between nominal circuit modes. Increasing it
hybridizes the normal modes more strongly: avoided crossings widen, the
computational eigenstates acquire more auxiliary-mode weight, and additional
driven paths become available. Reducing $s_{\rm mom}$ therefore suppresses a
dynamic source of parasitic Pauli terms, but it also changes the matrix element
through which the desired process is mediated.

The factor $s_{\rm pair}$ multiplies the modeled static pair interaction. This
term splits computational configurations according to their two-qubit parity
even with the pump turned off, and consequently appears primarily as
weight-two interactions in the Pauli table. The reference completion places
this static splitting far above the desired four-body scale. Lowering
$s_{\rm pair}$ is therefore a request for circuit-level compensation or
symmetry, not a numerical subtraction performed after the spectrum is found.

The equalization coordinate $e_\phi$ moves the four qubit participation phases
toward the common value $0.7$. These phases determine the relative signs with
which different virtual paths enter the cosine matrix element. Unequal phases
allow a process that should cancel by symmetry to survive, or cause the
components of the target process to add with unequal phases. Equalization
restores coherent addition of the requested channel and improves destructive
interference among symmetry-related parasitic paths.

Finally, $A_d$ is the dimensionless peak pump-flux excursion, expressed as a
phase in radians inside the Josephson cosine. In the weak-drive regime a larger
$A_d$ strengthens the resonant four-body matrix element. It simultaneously
increases ac-Stark shifts, off-resonant Floquet couplings, and mixing with
auxiliary configurations. Pump strength therefore moves the device along a
useful-interaction versus leakage tradeoff rather than supplying an
unconditionally beneficial scale factor.

\subsection{What the search reveals}

Screening uses eight lower-mode states, two upper-mode states, and
$\delta t=0.005\ns$. These levels include the dominant virtual lower-mode
configurations while making all 144 combinations affordable. Five points are
robustly accepted, one is unresolved at this resolution, and 138 are excluded.
Every robust survivor has $e_\phi=1$, showing that participation-phase
equalization is a structural requirement of this declared model rather than a
minor numerical improvement.

The nondominated choices reduce to two classes: $2\%$ momentum with $20\%$
static-pair retention, or $5\%$ momentum with complete pair cancellation. The
first preserves more of the static circuit interaction but demands stronger
suppression of momentum-mediated mixing; the second accepts more dynamic
mixing only if the modeled static pair term is cancelled. The balanced first
class is recomputed with 12 lower-mode states, two upper-mode states, and
$\delta t=0.00125\ns$. Enlarging the lower-mode basis tests whether omitted
virtual configurations appreciably shift the quasienergies or leakage. The
fine calculation gives
\begin{equation}
 |h_{\rm tar}|=80.916\khz,\quad
 h_{\rm driven}^{\max}/|h_{\rm tar}|=11.4\%,\quad
 h_{\rm static}^{\max}/|h_{\rm tar}|=0.641\%,\quad
 p_{\rm esc}=0.974\%.
 \label{sm:eq-cell-result}
\end{equation}
The neighboring $A_d=0.0048$ point gives $p_{\rm esc}=1.016\%$ and is
rejected. This nearby crossing of the declared leakage ceiling confirms the
physical tradeoff: the selected drive is the strongest sampled setting that
retains the four-qubit manifold within budget, rather than an arbitrary pump
chosen to reproduce the target coefficient. The source-informed reference
completion instead delivers $1.588\khz$ with a $1.834\,\mathrm{MHz}$ static
correlated channel. The search has therefore identified which spectral
features must change---phase alignment, momentum-mediated hybridization, and
static pair splitting---for the cell to become a plausible four-body coupler.
It does not by itself establish a complete code sector; that requires the
array-level calculation in Sec.~\ref{sm:sec-array}.

\section{Eight-qubit search and connected two-cell model}
\label{sm:sec-array}

\subsection{Uniform compiler target}
For each candidate setting, the solver recomputes the star and plaquette cells
using their actual four-frequency quartets. If delivered check coefficients
are $c_j$, the compiler requests
\begin{equation}
 c=\min_j|c_j|,\qquad H_0(c)=-c\sum_{j=1}^{8}S_j.
 \label{sm:eq-uniform-request}
\end{equation}
Every mismatch $c_j-c$ remains in $R$. This retains the proved uniform gap
$g_{\rm code}=4c$ instead of inferring a weighted-code gap.

\subsection{Connected pair correction}
An additive isolated-cell model misses interactions generated only when two
overlapping couplers operate simultaneously. Let $H_j$ be an isolated-cell
Hamiltonian and $H_{jk}$ the simultaneous result on the union of two cells.
After applying the same scalar and one-body conventions, define
\begin{equation}
 C_{jk}=H_{jk}-H_j-H_k,
 \qquad
 \effH^{(2)}=\sum_jH_j+\sum_{\langle j,k\rangle}C_{jk}.
 \label{sm:eq-connected-pair}
\end{equation}
The second sum contains all 24 cell pairs sharing a qubit. Pair propagation
acts on up to six qubits, with SQUID modes projected to their vacuum
expectation in this linked-cluster correction. The model includes connected
simultaneous-drive effects through cluster size two. It excludes three-cell
and higher connected processes and cross-cell virtual SQUID-mode excitation;
these are model boundaries, not terms covered by the numerical guard.

\begin{table}[t]
\caption{Connected-pair audit at the selected point. ``Max.'' is the largest
coefficient in $C_{jk}$; counts precede coalescence with other pairs.}
\label{sm:tab-pairs}
\scriptsize
\begin{ruledtabular}
\begin{tabular}{lclrr@{\quad}lclrr}
Cells & Shared & Norm & Max. & Terms & Cells & Shared & Norm & Max. & Terms\\
 & qubits & \multicolumn{2}{c}{(kHz)} & & & qubits & \multicolumn{2}{c}{(kHz)} & \\
\hline
s00--s01 & 1,2 & 8.122 & 3.302 & 935 & s00--s10 & 3,7 & 14.043 & 5.959 & 920\\
s00--p00 & 1,3 & 2.987 & 0.650 & 986 & s00--p01 & 2,3 & 3.923 & 1.118 & 947\\
s00--p10 & 1,7 & 3.068 & 1.027 & 942 & s00--p11 & 2,7 & 4.664 & 1.108 & 887\\
s01--s11 & 4,8 & 15.402 & 5.981 & 898 & s01--p00 & 1,4 & 3.778 & 2.402 & 976\\
s01--p01 & 2,4 & 9.555 & 3.216 & 897 & s01--p10 & 1,8 & 3.025 & 0.906 & 955\\
s01--p11 & 2,8 & 3.770 & 0.981 & 895 & s10--s11 & 5,6 & 12.832 & 5.198 & 922\\
s10--p00 & 3,5 & 7.665 & 2.944 & 951 & s10--p01 & 3,6 & 4.010 & 1.091 & 986\\
s10--p10 & 5,7 & 9.388 & 3.029 & 896 & s10--p11 & 6,7 & 4.812 & 2.856 & 967\\
s11--p00 & 4,5 & 5.725 & 2.837 & 898 & s11--p01 & 4,6 & 5.540 & 3.025 & 931\\
s11--p10 & 5,8 & 3.239 & 1.017 & 940 & s11--p11 & 6,8 & 3.388 & 0.673 & 944\\
p00--p01 & 3,4 & 14.198 & 6.143 & 921 & p00--p10 & 1,5 & 14.004 & 6.013 & 914\\
p01--p11 & 2,6 & 13.254 & 6.017 & 927 & p10--p11 & 7,8 & 8.073 & 4.205 & 924\\
\end{tabular}
\end{ruledtabular}
\end{table}

\subsection{Adaptive search}
The redesign varies nine coordinates over Table~\ref{sm:tab-array-domain}.
The search uses 96 seeded Latin-hypercube points plus four unique
anchor/amplitude controls, producing 100 global records. The best eight
feasible records each seed eight Gaussian local perturbations of fractional
radius $0.08$; with the retained anchor this gives 65 local records and 165
screened settings. The ten best feasible points plus the anchor are recomputed
at fine resolution.

\begin{table}[t]
\caption{Declared array redesign domain and selected fine candidate.}
\label{sm:tab-array-domain}
\begin{ruledtabular}
\begin{tabular}{p{0.42\textwidth}ccc}
Coordinate & Lower & Upper & Selected \\
\hline
High-frequency-qubit capacitance (fF) & 45 & 80 & 46.559 \\
Low-frequency-qubit capacitance (fF) & 95 & 150 & 132.847 \\
Inductance (nH) & 70 & 160 & 74.218 \\
Coupler $E_J/h$ (GHz) & 5 & 15 & 14.402 \\
Maximum drive phase (rad) & 0.0025 & 0.0080 & 0.005043 \\
Upper-mode frequency (GHz) & 35 & 100 & 46.355 \\
Retained momentum fraction & 0 & 0.02 & 0.01747 \\
Retained static-pair fraction & 0 & 0.10 & 0 \\
Participation-phase equalization & 0.90 & 1.00 & 0.95584 \\
\end{tabular}
\end{ruledtabular}
\end{table}

\begin{table}[tbp]
\centering
\small
\setlength{\tabcolsep}{4pt}
\renewcommand{\arraystretch}{1.15}
\caption{Testing conditions for the array experiment.
The preliminary requirements are checked numerically during
circuit selection. They are declared design and modeling
requirements, not fault-tolerance thresholds. The final
encoded-sector statement follows from the verified residual
bound, conditional on the declared model and numerical allowance.}
\label{sm:tab-array-testing-conditions}
\begin{ruledtabular}
\begin{tabular}{p{0.19\linewidth}p{0.36\linewidth}p{0.37\linewidth}}
Test & Required condition & Meaning \\ \hline

\multicolumn{3}{l}{\textit{Preliminary circuit requirements}} \\[2pt]

Computational leakage &
\(p_{\rm esc}^{(j)}<0.01\) for every coupling cell \(j\). &
After one drive period, the escape probability is below
\(1\%\) for every initial state in that cell's
computational subspace. \\[4pt]

Identification of qubit states &
Every assigned circuit eigenstate has squared overlap
at least \(0.90\) with its corresponding uncoupled
computational state. &
Ensures that the four-qubit states remain identifiable
after coupling to the SQUID modes. This is distinct
from leakage during driven evolution. \\[4pt]

Qubit--SQUID mixing &
The largest dimensionless mixing amplitude used in
the Schrieffer--Wolff reduction is at most \(0.15\). &
Limits mixing between qubit and SQUID excitations under
the chosen diagnostic; it does not itself bound the
error of the reduction. \\[4pt]

Hamiltonian extraction &
All extracted quasienergies remain strictly inside
the chosen principal Floquet interval
\((-1/(2T),1/(2T))\), in frequency units. &
Avoids the boundary where the selected matrix logarithm
is discontinuous. The drive period is
\(T=10\,\mathrm{ns}\). \\[4pt]

Qubit Josephson energies &
\(4.9\le E_{J,i}^{\rm bare}/h\le94\), in GHz,
for every qubit \(i\). &
Keeps the inferred bare qubit energies positive and
within the declared source-informed range.
These are not the coupler Josephson energies. \\[4pt]

Delivered check signs &
\(h_j<0\) for all eight stabilizer coefficients
in the adopted Pauli basis. &
Each requested stabilizer favors its satisfied
eigenvalue \(+1\). \\[4pt]

\multicolumn{3}{l}{\textit{Target construction and array tests}} \\[2pt]

Common target strength &
\(c=\min_j|h_j|\), with \(g_{\rm code}=4c\).
Every check-strength mismatch remains in \(R\). &
Compares the device with a uniform toric-code target,
rather than assuming that its eight delivered
strengths are identical. \\[4pt]

Driven and static residuals &
No separate percentage cap on the largest driven or
static correlated coefficient. All retained
mismatches enter \(R\). &
The array test assesses their combined action.
The single-cell limits of \(20\%\) and \(1\%\),
respectively, are not imposed here. \\[4pt]

Numerical array test &
\(2\normop{R_{\rm num}}<g_{\rm code}\), where
\(R_{\rm num}\) is the numerically assembled residual
including the modeled overlapping-cell contributions. &
Identifies candidates for subsequent verification.
This calculation alone is not the formally verified
result. \\[4pt]

Verified array test &
A verified bound \(\normop{R}\le\rho\), together with
the declared allowance \(\eta\), must satisfy
\(2(\rho+\eta)<g_{\rm code}\). &
Applies Theorem~\ref{thm:criterion} to establish four
isolated low-energy states and bound their subspace's
departure from the requested code space. \\

\end{tabular}
\end{ruledtabular}
\end{table}

Screening uses four lower-mode Fock states, two upper-mode states,
$\delta t=0.01\ns$, and a $10^{-12}\ghz$ Pauli threshold. Fine resolution
uses eight lower-mode states, two upper-mode states, $\delta t=0.0025\ns$,
and $10^{-13}\ghz$. A retained-candidate pair audit evaluates 31
candidate--resolution records: four pass after connected-pair correction,
including two fine candidates. The selected point is listed in
Table~\ref{sm:tab-array-domain}. Its delivered check magnitudes span
$165.742$--$175.942\khz$; the exact uniform target is
\begin{equation}
 c=165.741620811\khz,
 \qquad g_{\rm code}=4c=662.966483244\khz.
 \label{sm:eq-selected-target}
\end{equation}

\section{Residual anatomy and microscopic provenance}
\label{sm:sec-anatomy}

The final residual contains every retained difference between
$\effH^{(2)}$ and the uniform target: non-target terms from each isolated
cell, connected two-cell terms, and each delivered-check mismatch from $c$.
Coincident global Pauli words are summed before any norm is evaluated. The
exact inventory contains 8,561 nonzero words in the order
$(h_{00},h_{01},h_{10},h_{11},v_{00},v_{01},v_{10},v_{11})$.

As summarized graphically in Fig.~2(c) of the main text, the inventory spans
weights one through six. Two-qubit channels carry the largest aggregate
coefficient mass, $170.750\khz$, and the largest individual residual is the
$27.841\khz$ $ZZ$ interaction on qubits 1 and 3. Figure~2(d) identifies the
remaining leading words and their supports. The one-qubit $Y$ fields and the
weights three through six distinguish this connected-pair model from the
earlier additive model, whose compact inventory contained 1,708 non-target
words. That earlier inventory is not used in the certificate reported here.

The machine-readable record retains each isolated-cell table, all 24 pair
corrections, exact rational target and remainder words, and hashes tying the
generated Lean modules to their source. Coefficient anatomy is a diagnosis,
not a norm certificate: different Pauli directions cannot generally be
extremized by one state.

\section{Numerical guard and model boundary}
\label{sm:sec-guard}

The central Pauli table is a floating-point output at a selected truncation
and timestep. Lean cannot treat its diagonalized norm as an exact upper bound.
The witness therefore uses an outward central radius $\rho$, while $\eta$
covers three explicitly tested changes: the finite-Fock/base extraction
comparison, the connected-pair timestep comparison, and rationalization of
the pair table. Their respective contributions are
\begin{equation}
 \eta=0.539378843+1.745324516+0.0000041995
 =2.2847075585\khz.
 \label{sm:eq-guard}
\end{equation}

The guard does not enclose arbitrary fabrication error, calibration drift,
three-cell connected processes, cross-cell virtual SQUID-mode excitation, or
an omitted microscopic interaction. Finite-Fock comparisons are specified
convergence tests, not an interval proof over every cutoff. The certificate is
uniform over this guard and conditional on the connected-size-two model.

\section{Final exact certificate and central diagnostics}
\label{sm:sec-result}

For $g_{\rm code}/2=331.483241622\khz$, the coefficientwise bound is
$474.539596\khz$ centrally and $476.824304\khz$ after the guard, so it is
inconclusive. The exact anticommuting partition contains 795 groups, covers
every residual term once, and has largest group size 13. It gives
$246.358771\khz$ centrally and $248.643478\khz$ after the same guard, and
therefore passes. The dense witness gives the tightest certified values,
$142.000000\khz$ centrally and $144.284708\khz$ after the guard.

For the dense route, the numerical central norm is $141.821006\khz$ and the
exact proposed radius is $\rho=71/500000\ghz=142\khz$. The Cholesky factor
uses denominator $10^{15}$ and 32,896 nonzero Gaussian-integer entries. Lean
recomputes the defect; its maximum absolute row sum is 18,269,838,384,537
against slack numerator $10^{17}$, proving Eq.~\eqref{sm:eq-dense-witness}
without trusting the numerical Cholesky routine.

With $r=142+2.2847075585=144.2847075585\khz$, Lean verifies
\begin{equation}
 2r<g_{\rm code},
 \qquad g_{\rm code}-2r=374.397068127\khz,
 \label{sm:eq-final-corridor}
\end{equation}
so exactly four eigenvalues lie below $g_{\rm code}/2$, and
\begin{equation}
 \normop{P_{\rm low}-UP_0U^\dagger}
 \le\frac{r}{g_{\rm code}-r}<0.279.
 \label{sm:eq-final-projector}
\end{equation}
Direct diagonalization is not a theorem input but gives useful central
diagnostics: projector distance approximately $0.054$, low-band width
$3.927\khz$, and separation to the remaining spectrum $627.431\khz$.

\section{Formalization, agentic workflow, and trust boundary}
\label{sm:sec-formal}

The workflow separates candidate generation from mathematical verification. The agent performs parameter searches, extracts
Pauli coefficients, and proposes proofs and numerical witnesses. Import
scripts check the format and consistency of the numerical records before
their coefficients are represented as exact fractions. Lean then checks
proofs of the code properties, spectral perturbation results, Pauli-group
bounds, and matrix inequalities. Numerical estimates guide these
constructions but are not accepted as proofs of the corresponding bounds.

Some finite calculations use \texttt{native\_decide}, including the integer
inequalities underlying the dense residual bound and checks establishing
the ideal target's ground-space dimension and gap. These calculations rely
on Lean's compiler and runtime in addition to its proof-checking kernel.
They do not require treating the floating-point eigensolver or Cholesky
routine as correct: the proposed witness is checked using exact arithmetic.
The verification environment uses Lean 4.31.0 with fixed library revisions.
A dependency audit identifies the assumptions and native computations used
by the checked results.

Equal rank of $P_0$ and $P_{\rm low}$ guarantees a unitary $W$ satisfying
$P_{\rm low}=WP_0W^\dagger$. Consequently, the transported ideal logical
Pauli operators,
\begin{equation}
 \overline X_i^{\rm phys}=W\overline X_iW^\dagger,
 \qquad
 \overline Z_i^{\rm phys}=W\overline Z_iW^\dagger,
 \label{sm:eq-logical-transport}
\end{equation}
obey the same Pauli relations within the physical band. This algebraic
identification does not establish that $W$ is local or that the device
implements an error-recovery procedure.

Formal verification concerns the specified mathematical operators.
Circuit propagation, assignment of computational states, and convergence
tests remain numerical calculations. The extension of a residual bound
by the numerical allowance assumes that the discrepancy is no larger
than that allowance; its justification and limitations are described
in Sec.~\ref{sm:sec-guard}. Lean does not establish the accuracy of the
physical model, bound omitted interactions automatically, or verify the
behavior of a fabricated device.

\end{document}